\documentclass{aa}  

\usepackage{graphicx}
\usepackage{txfonts}
\usepackage{natbib}
\bibpunct{(}{)}{;}{a}{}{,}
\usepackage{enumitem}
\usepackage{epstopdf}
\usepackage[percent]{overpic}
\usepackage{color}
\usepackage{url}
\usepackage[]{xcolor}
\usepackage{tabularx,colortbl}
\usepackage{multirow}
\usepackage{comment}
\definecolor{white}{rgb}{1,1,1}
\definecolor{lightgrey1}{rgb}{0.95,0.95,0.95}
\definecolor{lightgrey2}{rgb}{0.8,0.8,0.8}
\definecolor{darkergrey}{rgb}{0.65,0.65,0.65}

\begin{document}
\authorrunning{Chatzistergos et al.}
\titlerunning{Irradiance reconstruction from Ca~II~K data}
\title{Reconstructing solar irradiance from historical Ca~II~K observations.}
\subtitle{I. Method and its validation.}
\author{Theodosios~Chatzistergos\inst{1,2}, 
        Natalie~A.~Krivova\inst{1}, 
        Ilaria~Ermolli\inst{2}\\
        Kok~Leng~Yeo\inst{1}, 
        Sudip~Mandal\inst{1}, 
        Sami~K.~Solanki\inst{1,3}, 
        Greg~Kopp\inst{4},
	    Jean-Marie~Malherbe\inst{5,6}
        }
\offprints{Theodosios Chatzistergos  \email{chatzistergos@mps.mpg.de}}
\institute{Max Planck Institute for Solar System Research, Justus-von-Liebig-Weg 3,	37077 G\"{o}ttingen, Germany 
        \and INAF Osservatorio Astronomico di Roma, Via Frascati 33, 00078 Monte Porzio Catone, Italy 
        \and School of Space Research, Kyung Hee University, Yongin, Gyeonggi 446-701, Republic of Korea
        \and Laboratory for Atmospheric and Space Physics, University of Colorado Boulder,
Boulder, Colorado, USA
\and LESIA, Observatoire de Paris, 92195 Meudon, France
\and PSL Research University, Paris, France }
\date{}

\abstract
{Knowledge of solar irradiance variability is critical to Earth's climate models and understanding the solar influence on Earth's climate.
Direct solar irradiance measurements are only available since 1978.
Reconstructions of past variability typically rely on sunspot data. 
However, sunspot  records provide only indirect information on the facular and network regions, which are decisive contributors to irradiance variability on timescales of the solar cycle and longer.
}
{Our ultimate goal is to
reconstruct past solar irradiance variations using historical full-disc Ca~II~K observations to describe the facular contribution independently of sunspot observations.
Here, we develop the method and test it extensively by using modern CCD-based Ca~II~K observations.
We also carry out initial tests on
two photographic archives.
}
{We employ carefully reduced and calibrated Ca~II~K images from 13 datasets, including some of the most prominent series, such as those from the Meudon, Mt Wilson, and Rome observatories.
We convert them to unsigned magnetic field maps and then use them as input to the adapted Spectral and Total Irradiance Reconstruction (SATIRE) model to reconstruct total solar irradiance variations over the period 1978--2019, for which direct irradiance measurements are available.
}
{The reconstructed irradiance from the analysed Ca~II~K archives agrees well with direct irradiance measurements and existing reconstructions.
The model also returns good results on data taken with different bandpasses and images with low spatial resolution.

Historical Ca~II~K archives suffer from numerous inconsistencies, but we show that these archives can still be used to reconstruct TSI with reasonable accuracy provided the observations are accurately processed and the effects of changes in instrumentation and instrumental parameters are identified and accounted for. 
The reconstructions are relatively insensitive  to the TSI reference record used to fix the single free parameter of the model.
Furthermore, even employment of
a series, itself reconstructed from Ca~II~K data, as a reference for further reconstructions returns nearly equally accurate results. This will enable the Ca~II~K archives without an overlap with direct irradiance measurements to be used to reconstruct past irradiance.}
{By using the unsigned magnetic maps of the Sun reconstructed from modern high-quality Ca~II~K observations as input into the SATIRE model, we can reconstruct solar irradiance variations nearly as accurately as from directly recorded magnetograms.  
Historical Ca~II~K observations can also be used for past irradiance reconstructions, but need additional care, e.g. to identify and account for discontinuities and changes in quality of the data with time.}


\maketitle

\section{Introduction}
\label{sec:intro}
\sloppy
\begin{table*}
	\caption{List of studies that reconstructed solar irradiance by using Ca~II~K observations.}
	\label{tab:irradiancereconstructions}     
	\centering                      
	\begin{tabular}{ll*{3}{c}}       
		\hline\hline                
		Study & \multicolumn{2}{c}{Ca~II~K data}& Years &Spectral range\\
		&  Type & Archive  &  &[nm]\\ 
		\hline\hline\\[-8pt]       
		\multicolumn{5}{c}{Empirical models}\\
		\hline
		\cite{oster_solar_1982}				&Disc-integrated plage area&	MW			&	1980	  & TSI\\ 	
		\cite{schatten_importance_1985} 	&Disc-integrated plage area&	MW			&  1978--1982&TSI\\ 			
		\cite{foukal_influence_1986}		&Disc-integrated plage area&	MW			&  1978--1982&TSI\\ 			
		\cite{foukal_magnetic_1988}			&Disc-integrated plage area& 	MW			&  1978--1984&TSI\\
		\cite{lean_variability_1983}		&Disc-integrated plage area&	MW			&  1969--1980&121.57\\
		\cite{lean_statistical_1987}		&Disc-integrated plage area&	MW			&  1978--1982&205\\
		\cite{pap_modelling_1991}           &Disc-integrated plage area&	BB			&  1986&125, TSI\\
		\cite{steinegger_energy_1996}		&Disc-integrated plage area&	SP			&  1980	  &TSI\\
		\cite{lean_magnetic_1998}			&Disc-integrated plage area&	BB			&  1991--1995&TSI, 120--420\\
		\cite{solanki_solar_1998}		&Disc-integrated plage area&	MW\tablefootmark{a}			&  1915--1984&TSI\\
		\cite{foukal_comparison_2002}		&Disc-integrated plage area&	MW, SP		&  1915--1999&TSI\\ 
		\cite{ambelu_estimation_2011}		&Disc-integrated index& 	MW, SP\tablefootmark{b}		    &  1915--2009&TSI\\	\cite{foukal_new_2012}				&Disc-integrated plage area& 	MW, SP		&  1915--1999&TSI, 130--240\\		
		\cite{morrill_solar_2011}			&Disc-integrated plage area&	BB			&  1991--1995&140--410\\
		\hline
		\cite{johannesson_reproduction_1995}&Disc-integrated intensity &	BB			&1991--1994&125\\
		\cite{chapman_variations_1996}		&Disc-integrated intensity &	SF			&1988--1992&TSI\\
		\cite{johannesson_10-year_1998}     &Disc-integrated intensity &	BB			&1987--1996&125\\
		\cite{preminger_photometric_2002}	&Disc-integrated intensity &	SF			& 1988--1996&TSI\\
		\cite{walton_contribution_2003}		&Disc-integrated intensity &	SF			& 1989--1996&TSI\\
		\cite{vogler_solar_2005}            &Disc-integrated intensity &    ML          & 1999&TSI\\
		\cite{chapman_comparison_2012}		&Disc-integrated intensity &	SF			& 2003--2010&TSI\\
		\cite{chapman_modeling_2013}		&Disc-integrated intensity &	SF			& 1988--2010&TSI\\
		\cite{puiu_modeling_2019}			&Disc-integrated intensity &   RP			& 2000--2019&TSI\\
		\cite{choudhary_variability_2020}	&Disc-integrated intensity &	SF			& 2003--2018&TSI\\
		\cite{chatzistergos_modelling_2020} &Disc-integrated intensity &   RP, SF		& 1996--2020&TSI\\
		\hline
		\cite{worden_far-ultraviolet_2001}  &Spatially resolved plage area&	SP			&1991--1996&120--170\\
		\cite{morrill_calculating_2005}		&Spatially resolved plage area&	BB			&1991--1995&276--288\\
		\cite{morrill_estimating_2011}		&Spatially resolved plage area&	MW			&1961--1981&276--288\\
		\hline
		\hline \\[-8pt]          
		\multicolumn{5}{c}{Semi-empirical models}\\
		\hline
		\cite{ermolli_modeling_2003}		&Spatially resolved plage area&	RP			& 	1996--2001&TSI\\
		\cite{penza_modeling_2003}			&Spatially resolved plage area&	RP			& 	1996--2001&TSI, 403, 501, 863\\	
		\cite{ermolli_recent_2011}			&Spatially resolved plage area&	RP			& 	1997--2010&TSI\\
		\cite{criscuoli_correlation_2018}	&Disc-integrated plage area   &	SF, ML		& 	1988--2015&115--310\\
		\cite{fontenla_bright_2018}			&Spatially resolved plage area&	Co, MD1, RP	& 	2002--2016&TSI,100--100,000\\
		This work							&Spatially resolved filling factors			&  13 archives  &	1978--2019&TSI, 115--160,000\\
		\hline
	\end{tabular}
	\tablefoot{Columns are: Bibliographic entry, type of input data and Ca~II~K archive(s) used, period and wavelength range of the reconstructed irradiance with Ca II K data. The studies are separated into four groups based on the method used for the irradiance reconstruction: empirical regression models using time-series of plage areas or Ca II K index;  empirical regression models using timeseries of disc-integrated intensities;  empirical models using time-series of plage areas and actual spectra; semi-empirical models using synthesized spectra from radiative transfer codes. See Table \ref{tab:observatories} for most of the archive abbreviations, while Co, ML, and SP refer to the Coimbra, Mauna Loa (centred at the core of the Ca~II~K line in contrast to the MLW data from the same observatory used here which were centred at the wing of the line), and Sacramento Peak solar observatories, respectively.
		\tablefoottext{a}{\citet{solanki_solar_1998} used the facular index compiled by \cite{fligge_long-term_1998} which included the MW plage areas by \cite{foukal_extension_1998}.
		}\tablefoottext{b}{\citet{ambelu_estimation_2011} used disc-integrated 1\AA~Ca~II~K index from SP and not the spectroheliograms.}}
\end{table*}

\newcounter{tableid}
\begin{table*}
	\caption{List of Ca~II~K datasets analysed in this study.}              
	\label{tab:observatories}      
	\centering 
	\small   
	\begin{tabular}{ll*{9}{c}}
		\hline\hline 
		Observatory & Acronym	   &Detector &Instrument&Period		  &Images&WL&SW   			          &Pixel scale			  	  & Ref.& Fig.\\
		&   		   &		 &			& 		      &      &[$\text{\AA}$]&[$\text{\AA}$]		  &[$"/$pixel] 			  	  &	\\
		\hline
		Baikal		 &Ba   &CCD& Filter& 2004--2019&789 &3933.7&1.2&2.7&\addtocounter{tableid}{1}\thetableid&\ref{fig:tsidaily_modern}\\[2pt]
		Big Bear  	 &BB   &CCD& Filter& 1982--2006  & 5026 &3933.7&3.2, 1.5\tablefootmark{a}&4.2, 2.4\tablefootmark{b}&\addtocounter{tableid}{1}\thetableid&\ref{fig:tsidaily_modernBBML}\\ [2pt]
		Brussels 	 &Br   &CCD& Filter& 2012--2019  &13248 &3933.7&2.7& 1.0&\addtocounter{tableid}{1}\thetableid&\ref{fig:tsidaily_modern}\\ [2pt]  
		Calern	     &CL   &CCD& Filter	& 2011--2019  & 1501&3933.7&7& 1.0&\addtocounter{tableid}{1}\thetableid&\ref{fig:tsidaily_modern}\\ [2pt]
		Kanzelh\"ohe &Ka   &CCD& Filter& 2012--2019  & 4146 &3933.7&3.0 &1.0&\addtocounter{tableid}{1}\thetableid&\ref{fig:tsidaily_modernBBML}\\ [2pt]
		Mauna Loa PSPT &MLW&CCD& Filter& 2004--2015  &9552 &3936.3&1.0& 1.0&\addtocounter{tableid}{1}\thetableid&\ref{fig:tsidaily_offbandmodern}\\ [2pt]
		Mees         &MS  &CCD& Filter 	& 1988--1998  & 1513 &3933.7&0.3\tablefootmark{c}& 5.5&\addtocounter{tableid}{1}\thetableid&\ref{fig:tsidaily_modern}\\ [2pt]
		Meudon	  	 &MD1 &Plate/CCD\tablefootmark{d}&SHG	    & 1978--2019 &11916 &3933.7&0.15, 0.09\tablefootmark{e}& 2.2, 1.5, 1.1\tablefootmark{f} &\addtocounter{tableid}{1}\thetableid&\ref{fig:tsidaily_historical}\\ 
		.... &MDW 		   &CCD&SHG	    & 2002--2018 &4481 &3932.3&0.15				  & 1.5 &\thetableid&\ref{fig:tsidaily_offbandmodern}\\  [2pt]
		Mount Wilson 		 &MW		   &Plate	 & SHG	    & 1978--1985  & 2413&3933.7&0.2 					  &2.9	  				  	  &\addtocounter{tableid}{1}\thetableid&\ref{fig:tsidaily_historical}\\ [2pt]
		Rome PSPT		     &RP   	   &CCD		 & Filter	& 1996--2019  & 3337 &3933.7&2.5 					  &2.0  	  &\addtocounter{tableid}{1}\thetableid&\ref{fig:pspt}\\ 
		San Fernando CFDT2	 &SF  		   &CCD	 	 & Filter 	& 1992--2013  & 4061 &3933.7&9  					  &2.6 						  &\addtocounter{tableid}{1}\thetableid&\ref{fig:tsidaily_modern}\\ [2pt] 
		Teide ChroTel		 &Te		   &CCD	 	 & Filter 	& 2009--2019  & 1708 &3933.7&0.3 					  &1.0 					 	  &\addtocounter{tableid}{1}\thetableid &\ref{fig:tsidaily_modern}\\  
		\hline
	\end{tabular}
	\tablefoot{Columns are: name of the observatory, abbreviation used in this study to refer to this dataset, type of detector, type of instrument, period of observations, total number of analysed images, central wavelength and spectral width of the spectrograph/filter, average pixel scale of the images, the bibliography entry, and the figure number in this paper showing the reconstructed TSI series with that dataset. \tablefoottext{a}{The two values correspond to the period before and after 10/09/1996, when the filter was changed.} \tablefoottext{b}{The two values correspond to the period before and after 08/11/1995, when the CCD camera was upgraded.}\tablefoottext{c}{This dataset comprises observations taken with an adjustable Lyot filter, the main settings for bandwidth and offset in the central wavelength should be 0.3~\AA~and 0~\AA, respectively. However, it is likely that it includes observations with the following settings too: [1.2,0]~\AA, [0.3,-0.6]~\AA, [0.6,-0.6]~\AA~for bandwidth and central wavelength combinations, respectively.}
		\tablefoottext{d}{The observations were stored on photographic plates up to 27/09/2002, while observations with a CCD camera started on 13/05/2002.}\tablefoottext{e}{The values refer to the periods [01/01/1978--14/06/2017] and [14/06/2017--31/12/2019].}\tablefoottext{f}{The values refer to the periods [01/01/1978--27/09/2002], [28/09/2002--14/06/2017], and [15/06/2017--31/12/2019].}}
	\tablebib{\addtocounter{tableid}{-\thetableid}
		(\addtocounter{tableid}{1}\thetableid) \citet{golovko_data_2002};
		(\addtocounter{tableid}{1}\thetableid) \citet{naqvi_big_2010};
		(\addtocounter{tableid}{1}\thetableid) \url{http://www.sidc.be/uset/}; 
		(\addtocounter{tableid}{1}\thetableid) \citet{meftah_solar_2018};
		(\addtocounter{tableid}{1}\thetableid) \citet{hirtenfellner-polanec_implementation_2011};
		(\addtocounter{tableid}{1}\thetableid) \citet{rast_latitudinal_2008};
		(\addtocounter{tableid}{1}\thetableid) \url{http://kopiko.ifa.hawaii.edu/KLine/index.shtml};
		(\addtocounter{tableid}{1}\thetableid) \citet{malherbe_new_2019}; 
		(\addtocounter{tableid}{1}\thetableid) \citet{lefebvre_solar_2005};
		(\addtocounter{tableid}{1}\thetableid) \citet{ermolli_photometric_2007};
		(\addtocounter{tableid}{1}\thetableid) \citet{chapman_solar_1997};
		(\addtocounter{tableid}{1}\thetableid) \citet{bethge_chromospheric_2011}.
	}
\end{table*}

\begin{table*}
	\caption{Comparison of TSI reconstructions from different Ca~II~K series (columns) to various TSI reference sets (rows).}
	\label{tab:tsibsat}     
	\centering                      
	\begin{tabular}{ll|*{1}{c|}*{3}{c}|*{3}{c}|*{2}{c}}       
		\hline\hline                
		&   &(1)&\multicolumn{3}{c|}{(2)}&
		\multicolumn{3}{c|}{(3)}&\multicolumn{2}{c}{(4)}\\[2pt]
		&	&RP&Te&Ba&MS&Br&CL&SF&MDW&MLW\\
		\hline
		& $B_{\mathrm{sat}}$ [G] &201 & 352 & 196 & 323 & 200 & 117 & 143 &  61 & 185 \\ 
		PMOD & RMS [Wm$^{-2}$] &\cellcolor{lightgrey1}{0.21} & \cellcolor{lightgrey1}{0.27} & \cellcolor{lightgrey1}{0.24} & \cellcolor{lightgrey2}{0.32} & \cellcolor{lightgrey1}{0.21} & \cellcolor{lightgrey1}{0.28} & \cellcolor{lightgrey1}{0.28} & \cellcolor{lightgrey1}{0.26} & \cellcolor{white}{0.18} \\ 
		& R &0.91 & 0.78 & 0.79 & 0.80 & 0.89 & 0.83 & 0.84 & 0.82 & 0.90 \\   \hline 
		& $B_{\mathrm{sat}}$ [G] &188 & 427 & 254 & 393 & 291 & 198 & 139 &  64 & 216 \\ 
		ACRIM & RMS [Wm$^{-2}$] &\cellcolor{lightgrey2}{0.35} & \cellcolor{lightgrey2}{0.33} & \cellcolor{lightgrey1}{0.28} & \cellcolor{darkergrey}{0.40} & \cellcolor{lightgrey1}{0.30} & \cellcolor{lightgrey2}{0.36} & \cellcolor{lightgrey2}{0.39} & \cellcolor{lightgrey2}{0.33} & \cellcolor{lightgrey1}{0.24} \\ 
		& R &0.81 & 0.48 & 0.58 & 0.63 & 0.53 & 0.58 & 0.73 & 0.63 & 0.65 \\   \hline 
		& $B_{\mathrm{sat}}$ [G] &205 & 361 & 198 & 339 & 212 & 122 & 143 &  63 & 189 \\ 
		RMIB & RMS [Wm$^{-2}$] &\cellcolor{lightgrey1}{0.23} & \cellcolor{lightgrey1}{0.27} & \cellcolor{lightgrey1}{0.23} & \cellcolor{lightgrey2}{0.33} & \cellcolor{white}{0.19} & \cellcolor{lightgrey1}{0.27} & \cellcolor{lightgrey1}{0.29} & \cellcolor{lightgrey1}{0.26} & \cellcolor{white}{0.18} \\ 
		& R &0.90 & 0.78 & 0.80 & 0.78 & 0.89 & 0.83 & 0.84 & 0.81 & 0.89 \\   \hline 
		& $B_{\mathrm{sat}}$ [G] &207 & 344 & 190 & 320 & 230 & 134 & 146 &  57 & 167 \\ 
		GEA18 & RMS [Wm$^{-2}$] &\cellcolor{lightgrey1}{0.23} & \cellcolor{lightgrey1}{0.26} & \cellcolor{lightgrey1}{0.29} & \cellcolor{lightgrey2}{0.33} & \cellcolor{lightgrey1}{0.21} & \cellcolor{lightgrey1}{0.29} & \cellcolor{lightgrey1}{0.28} & \cellcolor{lightgrey1}{0.26} & \cellcolor{white}{0.20} \\ 
		& R &0.90 & 0.81 & 0.74 & 0.78 & 0.86 & 0.81 & 0.83 & 0.83 & 0.90 \\   \hline 
		& $B_{\mathrm{sat}}$ [G] &200 & 353 & 197 & 476 & 200 & 117 & 141 &  61 & 185 \\ 
		SOHO/VIRGO & RMS [Wm$^{-2}$] &\cellcolor{lightgrey1}{0.22} & \cellcolor{lightgrey1}{0.28} & \cellcolor{lightgrey1}{0.25} & \cellcolor{white}{0.16} & \cellcolor{lightgrey1}{0.22} & \cellcolor{lightgrey1}{0.29} & \cellcolor{lightgrey1}{0.29} & \cellcolor{lightgrey1}{0.26} & \cellcolor{white}{0.19} \\ 
		& R &0.91 & 0.78 & 0.78 & 0.83 & 0.88 & 0.82 & 0.84 & 0.82 & 0.89 \\   \hline 
		& $B_{\mathrm{sat}}$ [G] &212 & 321 & 183 &  - & 200 & 110 & 143 &  60 & 157 \\ 
		SORCE/TIM & RMS [Wm$^{-2}$] &\cellcolor{white}{0.17} & \cellcolor{lightgrey1}{0.26} & \cellcolor{lightgrey1}{0.26} &  - & \cellcolor{white}{0.18} & \cellcolor{lightgrey1}{0.27} & \cellcolor{white}{0.20} & \cellcolor{lightgrey1}{0.24} & \cellcolor{white}{0.18} \\ 
		& R &0.91 & 0.82 & 0.78 &  - & 0.90 & 0.83 & 0.88 & 0.84 & 0.91 \\   \hline 
		& $B_{\mathrm{sat}}$ [G] &202 & 349 & 189 & 297 & 207 & 109 & 145 &  63 & 182 \\ 
		SATIRE-S & RMS [Wm$^{-2}$] &\cellcolor{white}{0.20} & \cellcolor{lightgrey1}{0.25} & \cellcolor{lightgrey1}{0.22} & \cellcolor{lightgrey2}{0.37} & \cellcolor{white}{0.18} & \cellcolor{lightgrey1}{0.29} & \cellcolor{lightgrey2}{0.30} & \cellcolor{lightgrey1}{0.26} & \cellcolor{white}{0.19} \\ 
		& R &0.92 & 0.80 & 0.81 & 0.74 & 0.91 & 0.82 & 0.81 & 0.81 & 0.89 \\   \hline 
		& $B_{\mathrm{sat}}$ [G] &238 & 493 & 260 & 396 & 328 & 203 & 165 &  70 & 213 \\ 
		SATIRE-T & RMS [Wm$^{-2}$] &\cellcolor{lightgrey2}{0.34} & \cellcolor{lightgrey2}{0.35} & \cellcolor{lightgrey2}{0.32} & \cellcolor{darkergrey}{0.46} & \cellcolor{lightgrey2}{0.32} & \cellcolor{lightgrey2}{0.36} & \cellcolor{lightgrey2}{0.39} & \cellcolor{lightgrey2}{0.35} & \cellcolor{lightgrey1}{0.27} \\ 
		& R &0.72 & 0.53 & 0.50 & 0.48 & 0.55 & 0.66 & 0.65 & 0.61 & 0.71 \\   \hline 
		& $B_{\mathrm{sat}}$ [G] &263 &  - & - & 332 &  - &  - & 189 &  43 & 142 \\ 
		SATIRE-T2 & RMS [Wm$^{-2}$] &\cellcolor{lightgrey2}{0.33} &  - & - & \cellcolor{darkergrey}{0.41} &  - &  - & \cellcolor{lightgrey2}{0.37} & \cellcolor{lightgrey2}{0.33} & \cellcolor{white}{0.19} \\ 
		& R &0.76 &  - & - & 0.70 &  - &  - & 0.68 & 0.77 & 0.77 \\   \hline   & $B_{\mathrm{sat}}$ [G] &188 & 310 & 164 & 297 & 174 & 103 & 137 &  50 & 152 \\ 
		EMPIRE & RMS [Wm$^{-2}$] &\cellcolor{lightgrey1}{0.25} & \cellcolor{lightgrey2}{0.30} & \cellcolor{lightgrey1}{0.30} & \cellcolor{lightgrey2}{0.37} & \cellcolor{lightgrey1}{0.24} & \cellcolor{lightgrey2}{0.33} & \cellcolor{lightgrey2}{0.31} & \cellcolor{lightgrey1}{0.27} & \cellcolor{lightgrey1}{0.21} \\ 
		& R &0.90 & 0.78 & 0.76 & 0.77 & 0.85 & 0.77 & 0.82 & 0.84 & 0.90 \\   \hline 
		& $B_{\mathrm{sat}}$ [G] &202 & 319 & 173 & 312 & 181 & 100 & 144 &  57 & 159 \\ 
		NRLTSI & RMS [Wm$^{-2}$] &\cellcolor{lightgrey1}{0.24} & \cellcolor{lightgrey1}{0.28} & \cellcolor{lightgrey1}{0.28} & \cellcolor{lightgrey2}{0.37} & \cellcolor{lightgrey1}{0.21} & \cellcolor{lightgrey2}{0.32} & \cellcolor{lightgrey2}{0.30} & \cellcolor{lightgrey1}{0.27} & \cellcolor{lightgrey1}{0.21} \\ 
		& R &0.89 & 0.79 & 0.76 & 0.76 & 0.89 & 0.78 & 0.82 & 0.82 & 0.89 \\   \hline 
		& $B_{\mathrm{sat}}$ [G] &213 & 331 & 181 & 332 & 185 & 107 & 153 &  57 & 162 \\ 
		NN-SIM & RMS [Wm$^{-2}$] &\cellcolor{lightgrey1}{0.23} & \cellcolor{lightgrey1}{0.30} & \cellcolor{lightgrey1}{0.29} & \cellcolor{lightgrey2}{0.35} & \cellcolor{lightgrey1}{0.23} & \cellcolor{lightgrey2}{0.34} & \cellcolor{lightgrey1}{0.30} & \cellcolor{lightgrey1}{0.27} & \cellcolor{lightgrey1}{0.21} \\ 
		& R &0.88 & 0.75 & 0.73 & 0.75 & 0.87 & 0.76 & 0.80 & 0.82 & 0.88 \\   \hline 
		& $B_{\mathrm{sat}}$ [G] &204 & 320 & 172 & 492 & 185 & 100 & 142 &  58 & 168 \\ 
		PSUM & RMS [Wm$^{-2}$] &\cellcolor{lightgrey1}{0.21} & \cellcolor{lightgrey1}{0.23} & \cellcolor{lightgrey1}{0.23} & \cellcolor{white}{0.19} & \cellcolor{white}{0.17} & \cellcolor{lightgrey1}{0.25} & \cellcolor{lightgrey1}{0.29} & \cellcolor{lightgrey1}{0.28} & \cellcolor{white}{0.19} \\ 
		& R &0.91 & 0.86 & 0.83 & 0.76 & 0.93 & 0.86 & 0.83 & 0.80 & 0.89 \\   \hline 
		& $B_{\mathrm{sat}}$ [G] &194 & 417 & 222 & 321 & 260 & 145 & 133 &  74 & 214 \\ 
		MD1 & RMS [Wm$^{-2}$] &\cellcolor{lightgrey1}{0.28} & \cellcolor{white}{0.19} & \cellcolor{white}{0.14} & \cellcolor{lightgrey2}{0.35} & \cellcolor{white}{0.13} & \cellcolor{lightgrey1}{0.23} & \cellcolor{lightgrey2}{0.35} & \cellcolor{lightgrey1}{0.21} & \cellcolor{white}{0.19} \\ 
		&RMS* [Wm$^{-2}$] &\cellcolor{lightgrey1}{0.22} & \cellcolor{lightgrey1}{0.28} & \cellcolor{lightgrey1}{0.25} & \cellcolor{lightgrey2}{0.33} & \cellcolor{lightgrey1}{0.24} & \cellcolor{lightgrey1}{0.29} & \cellcolor{lightgrey1}{0.29} & \cellcolor{lightgrey1}{0.27} & \cellcolor{white}{0.20} \\ 
		& R &0.87 & 0.86 & 0.92 & 0.80 & 0.94 & 0.86 & 0.80 & 0.85 & 0.84 \\   
		& $R$* &0.92 & 0.74 & 0.77 & 0.80 & 0.83 & 0.80 & 0.85 & 0.79 & 0.86 \\   \hline 
	\end{tabular}
	\tablefoot{The Ca II K series are divided into four groups: (1) RP (Sect. \ref{sec:results_pspt}); (2) datasets employing a narrower bandwidth than RP (Sect. \ref{sec:results_ccddata}); (3) datasets employing a broader bandwidth than RP (Sect. \ref{sec:results_ccddata}); (4) datasets that have off-band observations  (Sect. \ref{sec:results_ccddata}). 
		The bottom section of the table shows the results when the reference data set is, itself, the reconstruction from the MD1 Ca~II~K archive (see Sect. \ref{sec:sensitivity_reference} for details). In this case, we also show the results of a comparison of the respective reconstruction to the PMOD TSI composite, marked with an asterisk in the Table. The cells for the RMS differences are colour-coded from white to dark grey in the ranges [0,0.2], [0.2,0.3], [0.3,0.4], [0.4,0.5].}
\end{table*}

\begin{table}
	\caption{Comparison of TSI reconstructions from the resized RP Ca~II~K series to the PMOD TSI composite and the TSI reconstructions with original size RP data. 	}
	\label{tab:tsibsat_resized}     
	\centering                      
	\small
	\begin{tabular}{lccccc}       
		\hline\hline                
		Pixel scale&$B_{\mathrm{sat}}$&\multicolumn{2}{c}{RMS}&\multicolumn{2}{c}{$R$}\\
		$["/$pixel]& 	[G]&\multicolumn{2}{c}{[Wm$^{-2}$]}&&\\
		& 	 &RP&PMOD&RP&PMOD\\
		\hline
		2.0&201&-&0.214& &0.913\\ 
		2.2&201&0.029 & 0.213&0.998 & 0.914\\ 
		2.4&202&0.031 & 0.213&0.998 & 0.913\\ 
		2.8&203&0.033 & 0.214&0.998 & 0.913\\
		3.2&206&0.036 & 0.214&0.997 & 0.913\\ 
		3.9&208&0.041 & 0.212&0.997 & 0.915\\ 
		4.9&213&0.049 & 0.213&0.995 & 0.914\\ 
		6.5&223&0.060 & 0.214&0.993 & 0.914\\
		9.7&248&0.081 & 0.223&0.988 & 0.906\\\hline
	\end{tabular}
\end{table}

\begin{table}
	\caption{Variability in the measured and reconstructed TSI within one day for the first 7 days of June 2014.
	}
	\label{tab:tsibsatdaily}     
	\centering   
	\small                   
	\begin{tabular}{l|*{7}{c}}       
		\hline\hline                
		&	01&02&03&04&05&06&07\\
		\hline
		SOHO/VIRGO &   0.07&   0.07&   0.07&   0.06&   0.07&   0.07&   0.06\\
		SORCE/TIM &   0.11&   0.06&   0.07&   0.05&   0.06&   0.07&   0.09\\
		TCTE/TIM &   -&   -&   -&   -&   -& - &   0.11\\
		Ba &  -&   0.08&   0.13&   0.09&   0.10&  -&   -\\
		Br &   0.04&   0.10&   -&   -&   0.03&   0.16&   0.12\\
		CL &   -&  -&   -&   0.09&   0.10&   0.18&   -\\
		Ka &   0.14&   0.07&   0.10&   0.07&   0.11&   0.09&   0.07\\
		Te &   0.05&   0.15&   0.12&   0.13&   0.09&   0.18&   0.16\\
		\hline
	\end{tabular}
	\tablefoot{The values are the standard deviations of all reconstructed TSI values within each given day (numbered in the top row) after subtracting a series resulting from 30 minute smoothing of the SOHO/VIRGO values in order to take into account the evolution of solar activity.}
\end{table}

\begin{table}
	\caption{Comparison of TSI reconstructions from the RP and two photographic Ca~II~K archives to the PMOD TSI composite and SATIRE-S, SATIRE-T, and SATIRE-T2 models.}
	\label{tab:tsibsathistorical}     
	\centering   
	\small                   
	\begin{tabular}{l|*{4}{c}}       
		\hline\hline                
		&	PMOD&SATIRE-S&SATIRE-T&SATIRE-T2\\
		\hline
		RP & 0.24 (0.91) & 0.24 (0.92) & 0.40 (0.76) & 0.42 (0.76) \\  PMOD &  - & 0.17 (0.98) & 0.33 (0.86) & 0.38 (0.85) \\ \hline
		SF & 0.30 (0.84) & 0.31 (0.83) & 0.42 (0.69) & 0.43 (0.70) \\  PMOD &  - & 0.18 (0.95) & 0.31 (0.83) & 0.36 (0.81) \\ \hline
		MD1 & 0.41 (0.77) & 0.45 (0.72) & 0.56 (0.58) & 0.56 (0.61) \\  PMOD &  - & 0.24 (0.91) & 0.39 (0.77) & 0.36 (0.83) \\ \hline
		MW & 0.43 (0.75) & 0.41 (0.74) & 0.56 (0.58) & 0.54 (0.67) \\  PMOD &  - & 0.33 (0.86) & 0.47 (0.74) & 0.40 (0.83) \\ \hline
		\hline
	\end{tabular}
	\tablefoot{The values are RMS differences in Wm$^{-2}$ followed by $R$ in parentheses. The 2nd line of each section of the table compares SATIRE-S/T/T2 modeled series to PMOD TSI over the same dates as used in the Ca II K archive in question. }
\end{table}

\begin{figure}
	\centering
	\includegraphics[width=1\linewidth]{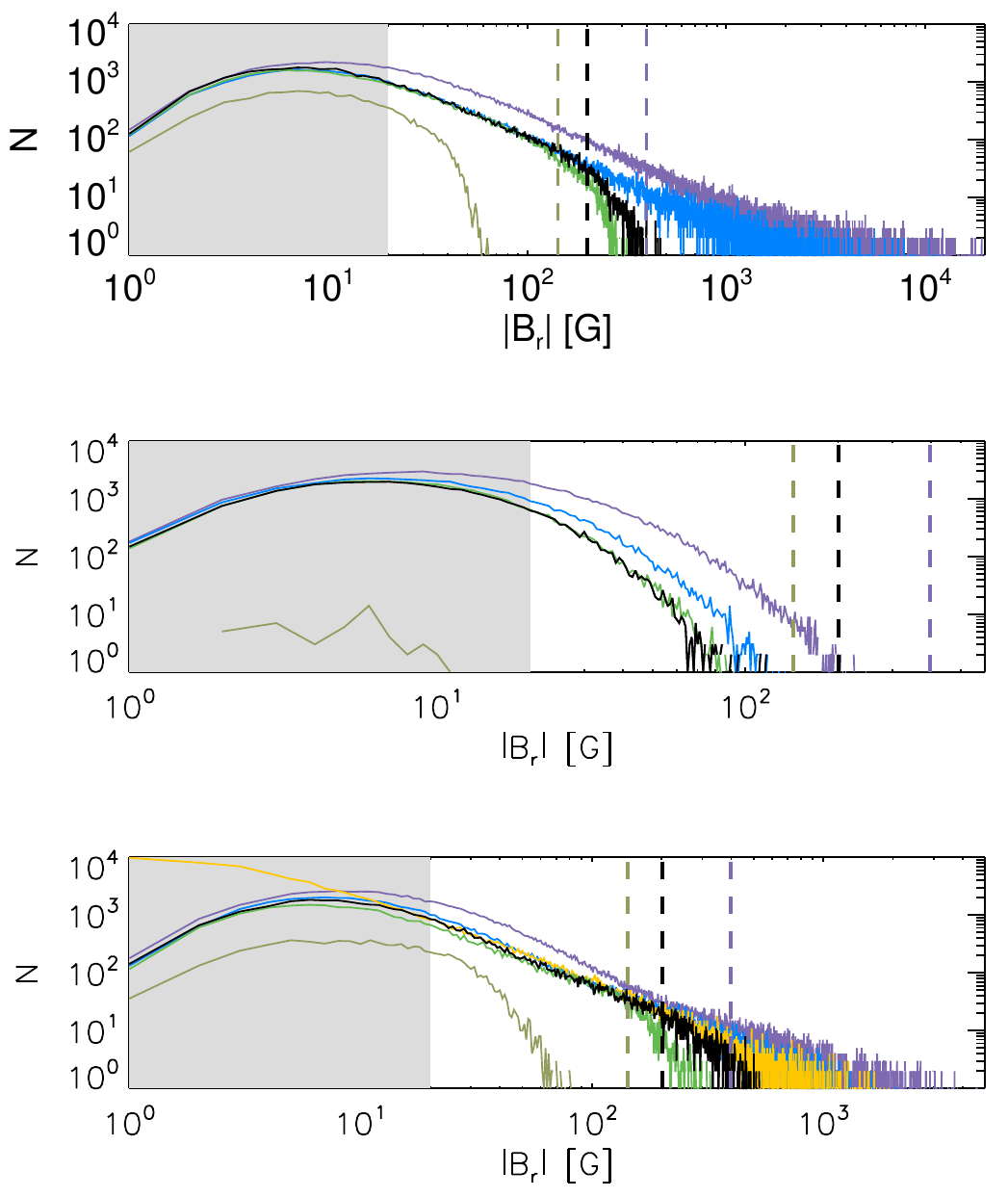}
	\caption{	Histograms of the unsigned radial magnetic flux density $|B_r|$ in magnetograms reconstructed from Ca~II~K observations taken on 11 August 2001 (top), 16 July 2009 (middle), and 10 May 2012 (bottom).
		The observations used are from RP (black), MD1 (blue/purple for reconstructions with/without the normalisation of the contrast values to RP, see Sect. \ref{sec:data}),  SF (dark/light green for reconstructions with/without the normalisation), and SDO/HMI (orange, shown only in the bottom panel as these data are not available prior to 2010) archives. 
		The histograms are produced for bins of 1 G. The vertical dashed lines mark the adopted $B_{\mathrm{sat}}$ value for each archive when using PMOD TSI as the reference, while the grey shaded surface denotes the 3-$\sigma$ noise level ($\sim$20 G) of the SDO/HMI magnetograms.}
	\label{fig:histograms}
\end{figure}

\begin{figure*}
	\centering
	\includegraphics[width=1\linewidth]{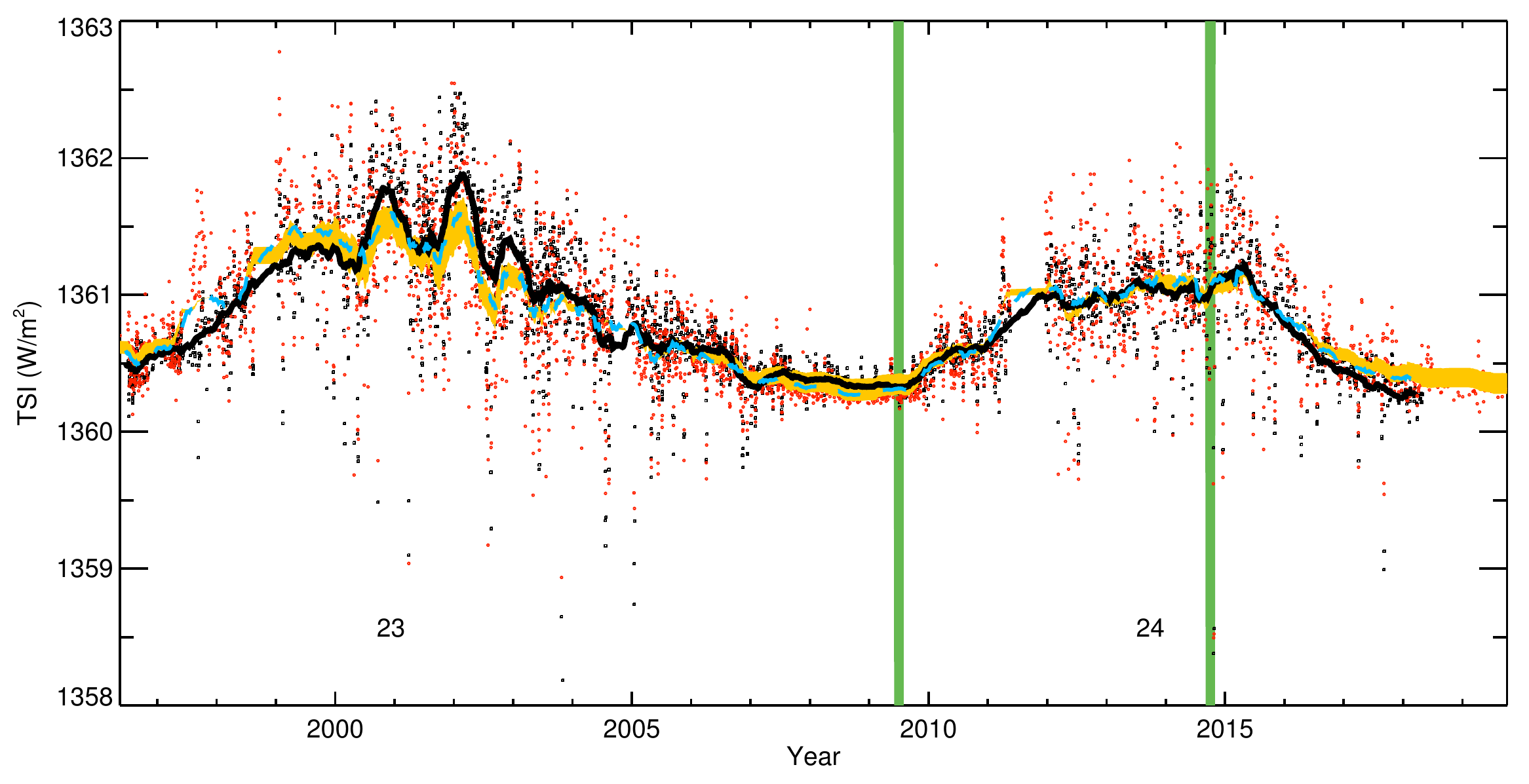}
	\caption{TSI reconstructed with RP Ca~II~K data (red circles for daily values and light blue dashed line for 81-day running mean values) using the PMOD TSI series as the reference as a function of time. Also shown is the PMOD TSI composite (black squares for daily values and black solid line for 81-day running mean values). The orange shading shows the range of the TSI reconstructions (81-day running means) obtained with the minimum and maximum $B_\mathrm{sat}$ value determined from all reference TSI series considered here (see Table~\ref{tab:tsibsat}). The numbers underneath the curves denote the solar cycle numbers, while the green vertical bars mark the intervals shown in Fig.~\ref{fig:psptdaily}. The running means are produced by considering only the days on which both series are available.}
	\label{fig:pspt}
\end{figure*}

\begin{figure*}
	\centering
	\includegraphics[width=1\linewidth]{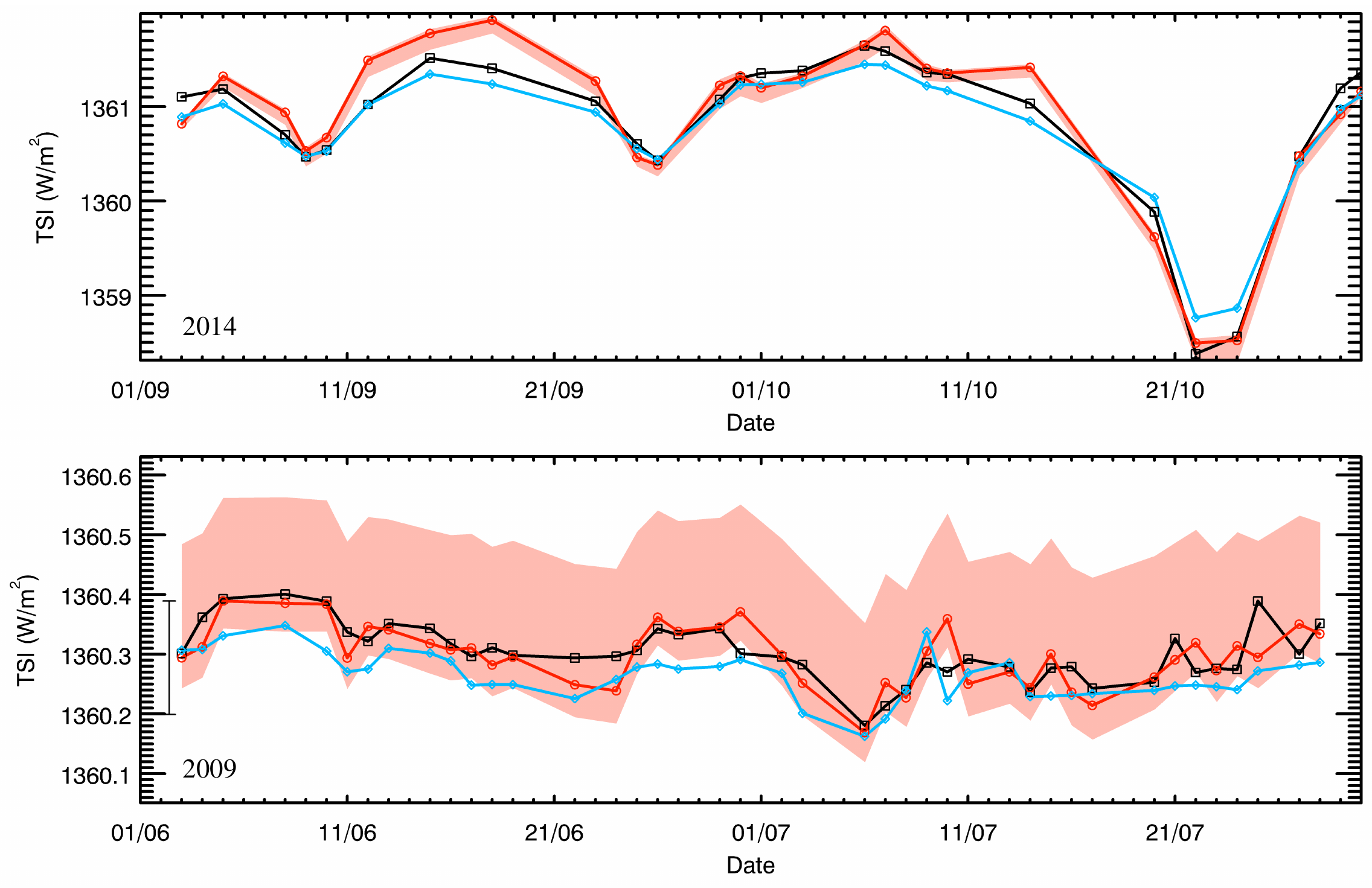}
	\caption{TSI reconstructed from RP Ca~II~K data (red circles for daily values connected by the red line) during 2-month periods September--October 2014 
		(close to activity maximum, top panel) and June--July 2009 (around activity minimum, bottom panel). Also shown are the PMOD TSI composite (black squares and black curve) and the SATIRE-S TSI series (light blue diamonds and line). 
		All series are only shown on the days when RP observations are available.
		The red shaded area shows the entire range of TSI obtained from RP Ca~II~K 
		data using different reference series. 
		Note that the range of TSI values shown in the two panels differs. 
		To ease the comparison, the error bar in the lower panel marks the maximum range of the red shaded surface over the 2-month interval in 2014 shown in the top panel.}
	\label{fig:psptdaily}
\end{figure*}

\begin{figure}
	\centering
	\includegraphics[width=1\linewidth]{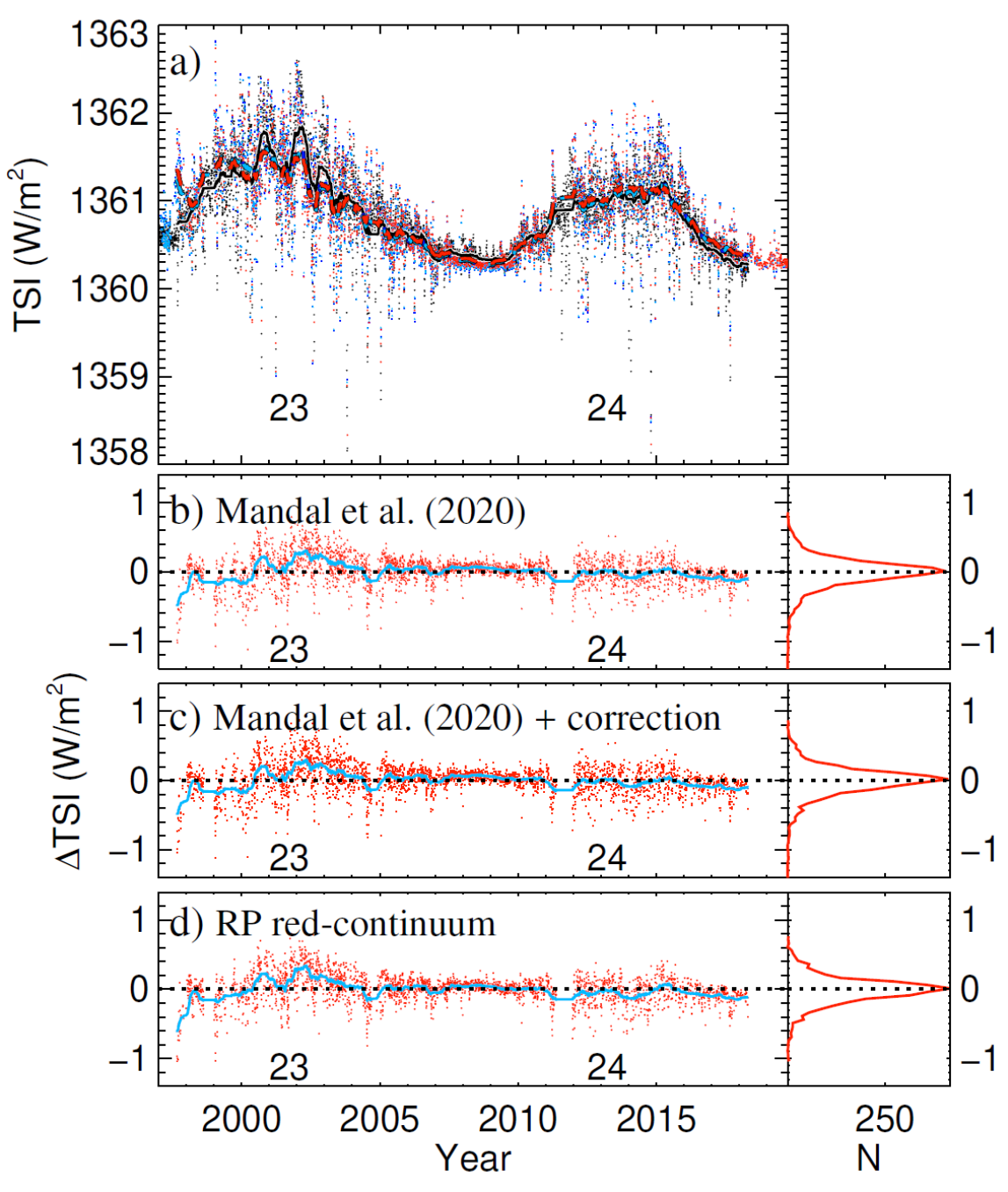}
	\caption{TSI reconstructed from RP Ca~II~K data as a function of time (a) and difference of PMOD TSI composite to the TSI reconstructions (b--d) when using different approaches to account for sunspots: using the \cite{mandal_sunspot_2020} series to get the sunspot filling factors (light blue in panel a and red in panel b); using the \cite{mandal_sunspot_2020} series to get the sunspot filling factors which are then subtracted from facular filling factors (blue in panel a and red in panel c); using full-disc Rome/PSPT red continuum data (red in panel a and d) to get the sunspot filling factors (see Sect. \ref{sec:sunspotcontribution} for more information). The differences are shown only for the common days in all series. Also shown in panel a is the PMOD TSI composite (black). Thick lines show 81-day running mean values. The right parts of the lower panels show the distributions of the differences in bins of 0.05 Wm$^{-2}$.See Fig. \ref{fig:pspt11scatter} for the RMS differences between the various series as well as the correlation coefficient, $R$.}
	\label{fig:psptdifsunspotdif}
\end{figure}

\begin{figure*}
	\centering
	\includegraphics[width=1\linewidth]{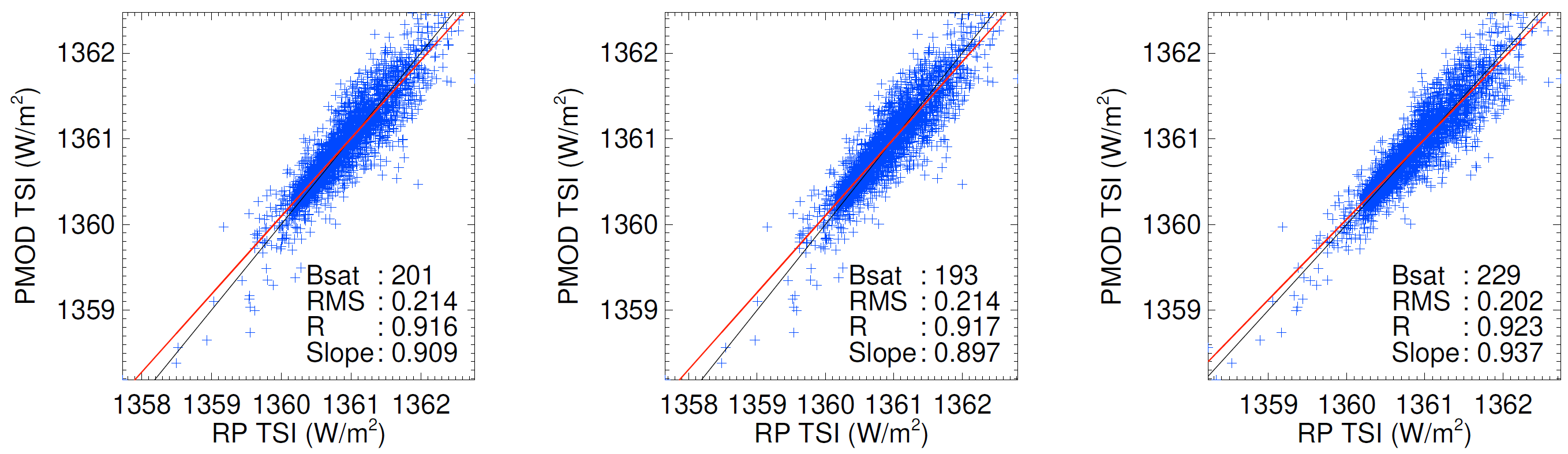}
	\caption{TSI reconstructed from RP Ca~II~K images against the PMOD TSI composite. \textit{Left:} sunspot information taken from \cite{mandal_sunspot_2020} without any correction to the faculae filling factors due to sunspots; \textit{Middle:} sunspot information taken from \cite{mandal_sunspot_2020} and the sunspot filling factors are subtracted from the facular ones; \textit{Right:} the sunspot information is derived from the full-disc RP red-continuum images. The red lines show linear fits to the data, while the black lines have a slope of unity. Also listed in each panel are the B$_{\mathrm{sat}}$ (in G), RMS difference (in W/m$^{2}$), the linear correlation coefficient, and the slope of the linear fit.}
	\label{fig:pspt11scatter}
\end{figure*}

\begin{figure*}
	\centering
	\includegraphics[width=1\linewidth]{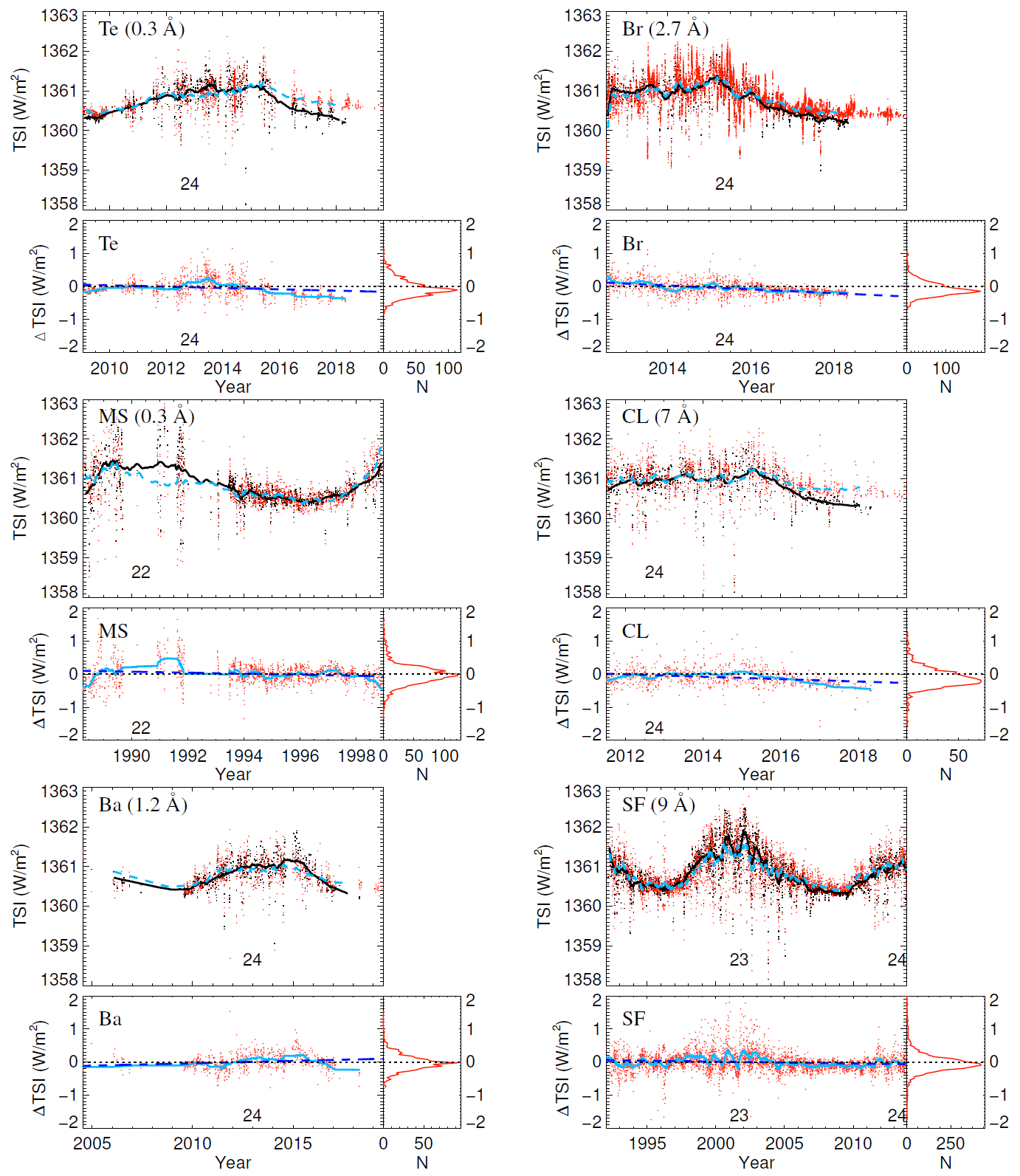}
	\caption{\textit{Upper part of each panel: }
		TSI reconstructions from Ca~II~K observations taken with filters that are narrower (left column) or broader (right column) than RP.
		Daily values from the Te, Br, MS, CL, Ba, and SF datasets are shown as red dots and the corresponding 81-day running means as dashed blue lines.
		For comparison, the PMOD irradiance composite series is overplotted (black dots for daily values and the solid black line for the 81-day running means). The running means have been computed considering only days on which both the reconstructions and the PMOD series were available. 
		\textit{Lower left part of each panel: } Difference between the PMOD  composite and the reconstructed TSI (red dots for daily values and solid light blue for 81-day running means). The dashed blue line is a linear fit to the residuals.
		The horizontal black dotted line marks the zero difference. 
		\textit{Lower right part of each panel:} distribution of the residuals in bins of 0.05 Wm$^{-2}$.
		The numbers at the bottom of each panel indicate the solar cycle number and are placed roughly around the maximum of the corresponding cycle. 	}
	\label{fig:tsidaily_modern}
\end{figure*}

\begin{figure*}
	\centering
	\includegraphics[width=1\linewidth]{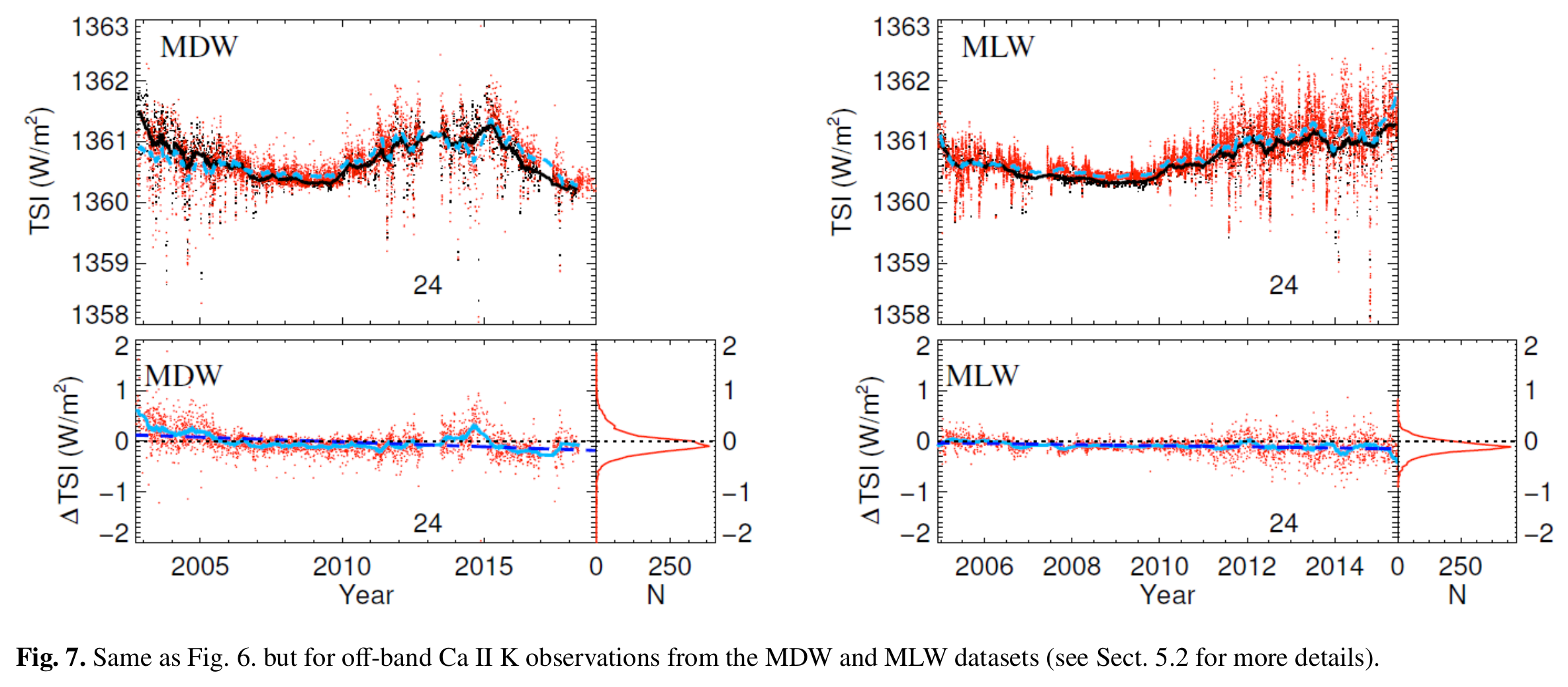}
	\caption{Same as Fig. \ref{fig:tsidaily_modern}. but for off-band Ca~II~K observations from the MDW and MLW datasets (see Sect.~\ref{sec:results_ccddata} for more details).} 
	\label{fig:tsidaily_offbandmodern}
\end{figure*}

\begin{figure*}
	\centering
	\includegraphics[width=1\linewidth]{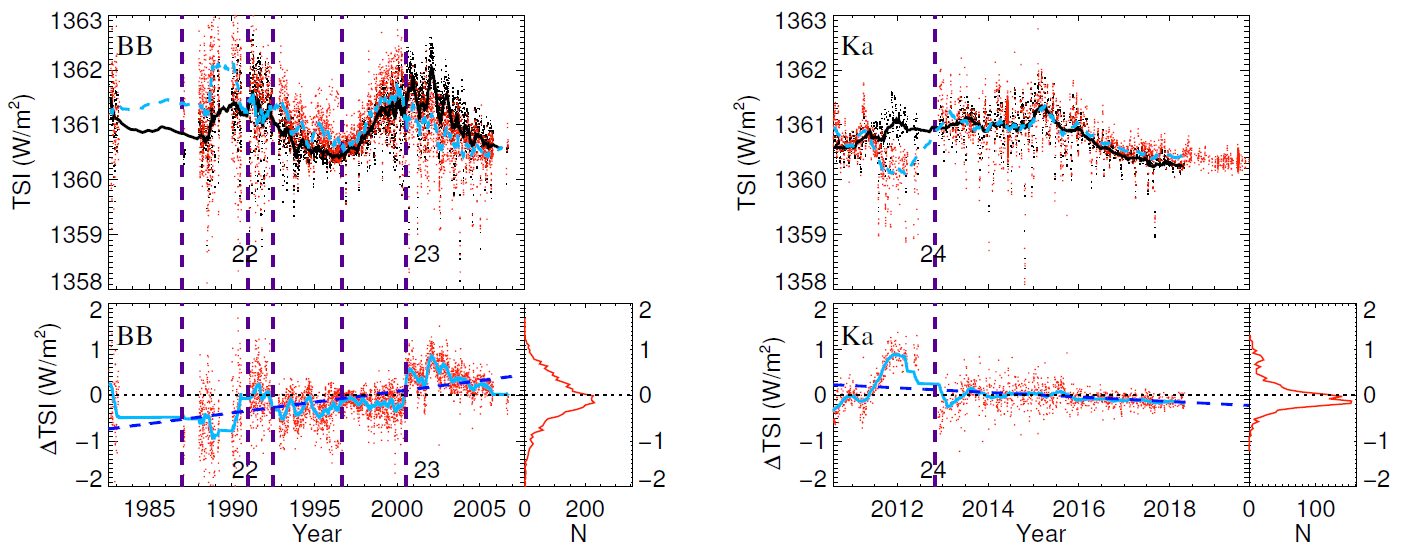}
	\caption{
		Top panels: Daily TSI from the PMOD composite (black) and reconstructed (red) from Ca~II~K observations from the BB (left) and  Ka datasets (right). Also shown are 81-day running mean values  (dashed ciel lines for the reconstructions and black solid  lines for the PMOD composite) but only for the days, when both PMOD and the respective reconstruction are available. 
		Bottom panels: The difference between the PMOD TSI series and the reconstructed TSI.
		The ciel line shows the 81-day running mean, while the dashed blue line is  a linear fit to the residuals. The horizontal dotted black lines denote zero difference. The numbers at the bottom of the panels indicate the solar cycle number and are plotted at roughly the maximum of each cycle. Periods with known instrumental or observational changes are marked with vertical dashed purple lines 
		(see Sec.  \ref{sec:results_ccddata} for more details).}
	\label{fig:tsidaily_modernBBML}
\end{figure*}

\begin{figure*}
	\centering
	\includegraphics[width=1\linewidth]{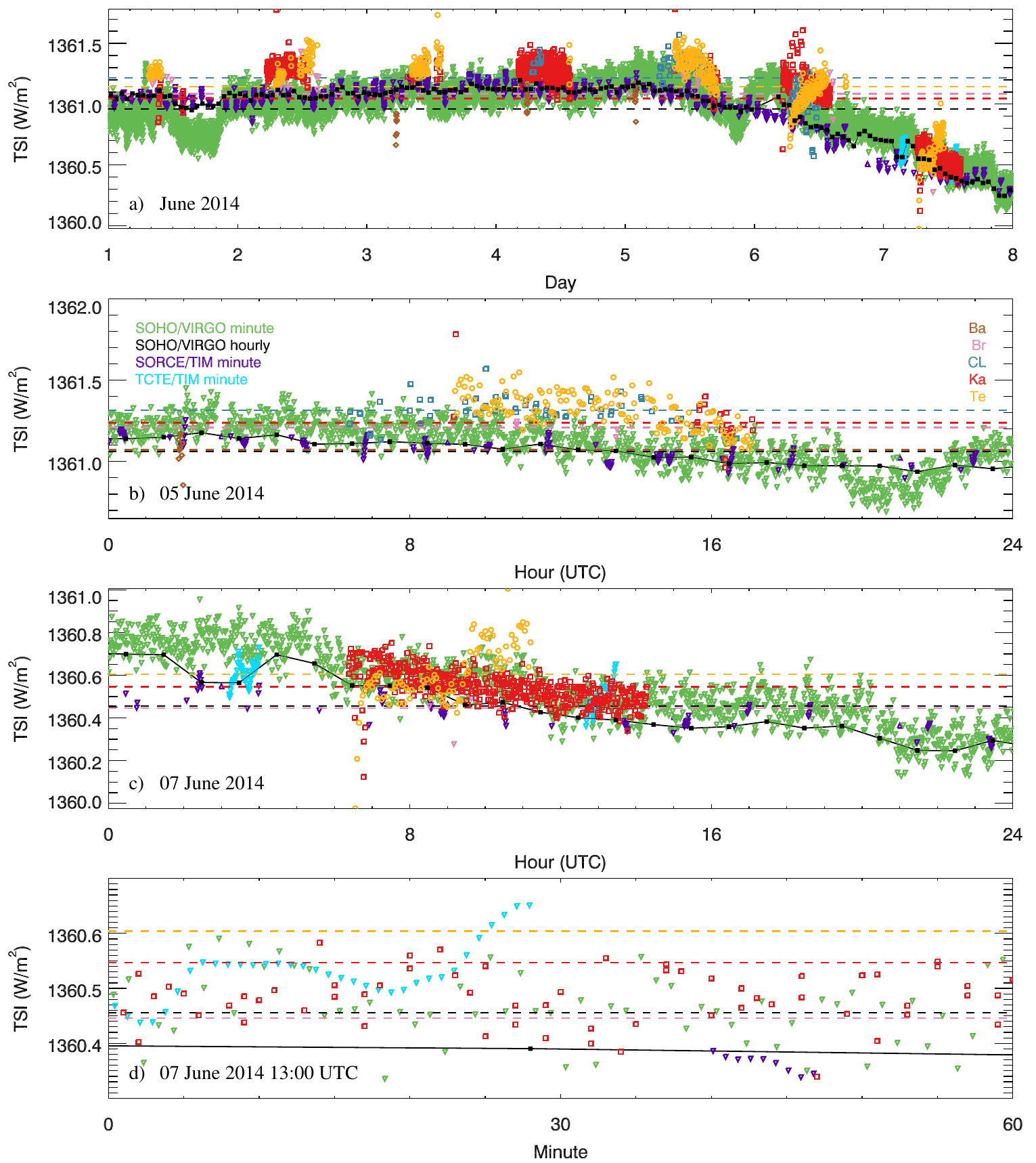}
	\caption{Reconstructed TSI from Ca II K observations using SOHO/VIRGO data (daily values) as the reference over the course of the first week of June 2014 (panel a), on 05 June 2014 (panel b),  07 June 2014 (panel c), and between 13:00 and 14:00 UTC on 07 June 2014 (panel d). The reconstructed TSI is based on the images from Ba (brown rhombuses), Br (pink downward triangles), CL (blue empty squares), Ka (red empty squares), and Te (orange circles). Also shown are the TSI measurements by SOHO/VIRGO (black filled squares connected by the black solid line for hourly values and green downward triangles for data with the cadence of 60~s), SORCE/TIM (purple downward triangles; cadence of 50 s), and TCTE/TIM (ciel downward triangles; cadence of 50 s).
		In contrast to all other figures where TSI series reconstructed from Ca~II~K data are shown as daily means over all individual images on that day, here we show the results for each individual image. The horizontal dashed lines mark the mean TSI value of each corresponding series over the specified period, except for panel d, in which the horizontal lines mark the daily mean value.}
	\label{fig:201406}
\end{figure*}

\begin{figure*}
	\centering
	\includegraphics[width=1\linewidth]{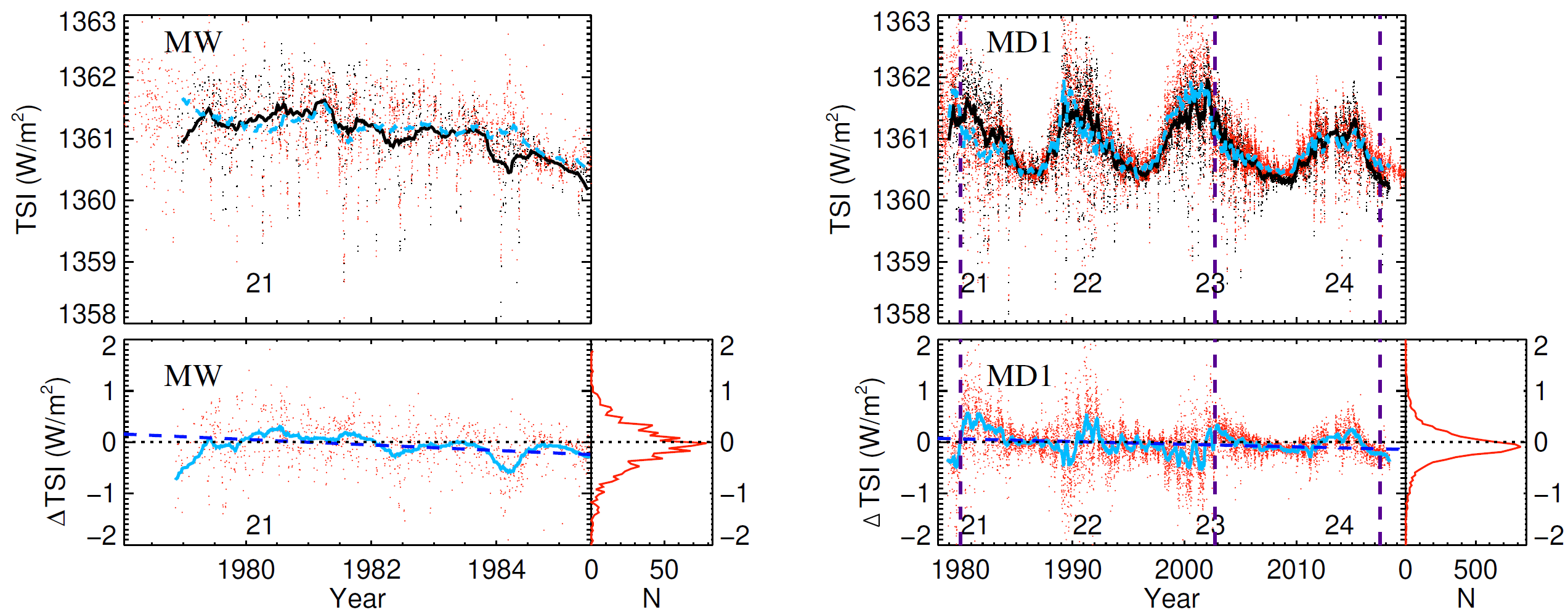}
	\caption{\textit{Upper part of each panel: }
		TSI reconstructions from photographic Ca~II~K observations.
		Daily values from the MW and MD1 datasets are shown as red dots and the corresponding 81-day running means as dashed blue lines.
		For comparison, the PMOD irradiance composite series is overplotted (black dots for daily values and the solid black line for the 81-day running means), which was used as the reference for the reconstructed series from MD1 Ca~II~K data. 
		The running means have been computed considering only days on which both the reconstructions and the PMOD series were available. 
		\textit{Lower left part of each panel: } Difference between the PMOD  composite and the reconstructed TSI (red dots for daily values and solid light blue for 81-day running means). 
		The dashed blue line is a linear fit to the residuals.
		The horizontal black dotted line marks the zero difference. 
		\textit{Lower right part of each panel:} distribution of the residuals in bins of 0.05 Wm$^{-2}$.
		The numbers in the bottom part of each panel indicate the solar cycle number, which are placed roughly at the maximum of each cycle.
	}
	\label{fig:tsidaily_historical}
\end{figure*}

\begin{figure*}
	\centering
	\includegraphics[width=1\linewidth]{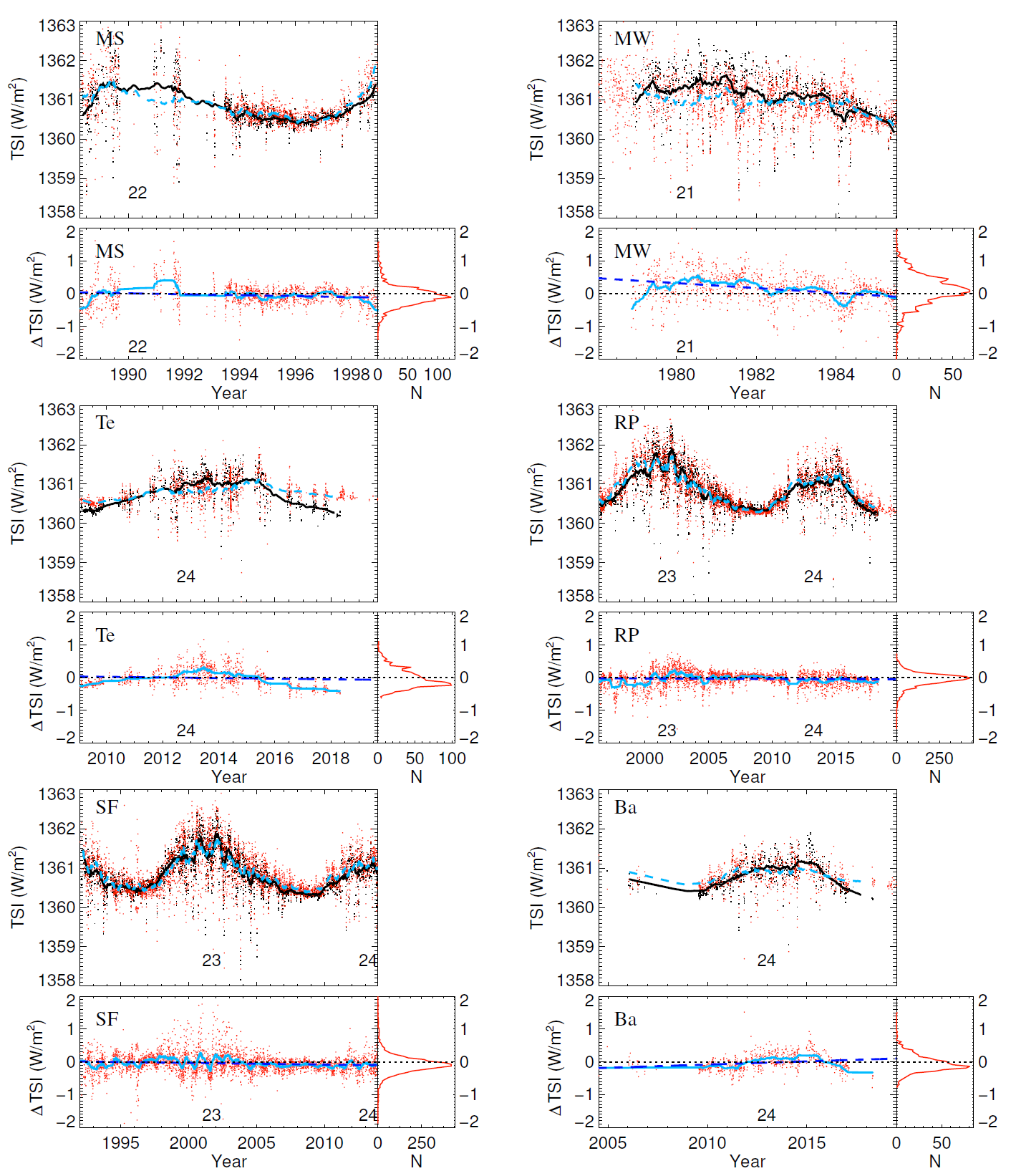}
	\caption{\textit{Upper part of each panel: }
		TSI reconstructions from Ca~II~K observations using the reconstructed TSI series from MD1 Ca~II~K data as the reference.
		Daily values from the MS, MW, Te, RP, SF, and Ba datasets are shown as red dots and the corresponding 81-day running means as dashed blue lines.
		For comparison, the PMOD irradiance composite series is overplotted (black dots for daily values and the solid black line for the 81-day running means), which was used as the reference for the reconstructed series from MD1 Ca~II~K data. 
		The running means have been computed considering only days on which both the reconstructions and the PMOD series were available. 
		\textit{Lower left part of each panel: } Difference between the PMOD  composite and the reconstructed TSI (red dots for daily values and solid light blue for 81-day running means). 
		The dashed blue line is a linear fit to the residuals.
		The horizontal black dotted line marks the zero difference. 
		\textit{Lower right part of each panel:} distribution of the residuals in bins of 0.05 Wm$^{-2}$.
		The numbers in the bottom part of each panel indicate the solar cycle number, which are placed roughly at the maximum of each cycle.
	}
	\label{fig:tsidaily_meudonref}
\end{figure*}

The Sun is the dominant energy source to Earth's system  \citep{kren_where_2017} and is thus one of the main natural drivers of Earth's climate.
However, the mechanisms of this forcing are not yet fully understood, due to the complex dynamics of the climate system and scarcity of appropriate solar data \citep[e.g.,][]{haigh_sun_2007,gray_solar_2010,ermolli_recent_2013,solanki_solar_2013-1,krivova_solar_2018}.

Climate models require information on past changes in the solar irradiance \citep{intergovernmental_panel_on_climate_change_climate_2013,intergovernmental_panel_on_climate_change_climate_2021,matthes_solar_2017}, 
which is the solar radiative energy flux per unit area as measured at the top of the Earth's atmosphere at a mean Sun-Earth distance of one astronomical unit.
The solar irradiance integrated over the entire spectrum,  called the total solar irradiance (TSI), describes the total solar energy input and has been measured nearly continuously since 1978 by various radiometers from space \citep[e.g.][]{hickey_initial_1980,willson_solar_1988,frohlich_-flight_1997,kopp_total_2005,schmutz_total_2013,pilewskie_tsis-1_2018}. 
These measurements show that TSI varies at all discernible timescales from minutes to decades, most distinctive being a clear $\sim$0.1\% change in phase with the solar cycle \citep{frohlich_total_2013,kopp_magnitudes_2016}.
There is also indirect evidence for longer-term irradiance changes, which are of primary interest to climate studies. 
Since no irradiance measurements are available for the period prior to 1978, models are used to reconstruct past variations.

It has been demonstrated that the main driver of the irradiance variability on timescales of days to decades is the perpetual competition between
the dark and bright magnetic regions emerging through the solar surface \citep[see e.g.][and references therein]{foukal_effect_1977,yeo_solar_2017,shapiro_nature_2017}, i.e. the dark sunspots and bright faculae and network (hereafter collectively referred to as faculae) seen in white light in the photosphere. 
Thus, models have been developed that reconstruct solar irradiance changes by accounting for sunspot darkening and facular brightening.
Most successful models reproduce over 90\% of the measured TSI variability \citep[e.g.,][]{krivova_reconstruction_2003,chapman_modeling_2013,yeo_reconstruction_2014,yeo_empire_2017,yeo_solar_2017,dasi-espuig_reconstruction_2016,lean_estimating_2018,wu_solar_2018-2,chatzistergos_modelling_2020}. 
The success of these models suggests that it is possible to reconstruct past irradiance variability, provided the contributions by spots and faculae are known.

The compiled series of counts of sunspots (the sunspot number) and their groups (the group sunspot number) go back to 1700 
and 1609, respectively, and have been most widely used for irradiance reconstructions extending back to the Maunder Minimum.
As with most historical data, they are not free of problems.
In particular, the cross-calibration of records by individual observers has recently been debated \citep{clette_new_2016-1,clette_new_2018,lockwood_tests_2016-2,usoskin_dependence_2016}.
This led to a development of new techniques \citep{usoskin_new_2016,willamo_updated_2017,chatzistergos_new_2017}, re-analysis of some data, as well as recovery of additional sources \citep[e.g.][]{arlt_sunspot_2013,arlt_sunspot_2016,vaquero_revised_2016,carrasco_sunspot_2018-1,carrasco_two_2019,carrasco_note_2021,carrasco_forgotten_2021,hayakawa_thaddaus_2020,hayakawa_sunspot_2020,hayakawa_daniel_2021,vokhmyanin_sunspot_2020}, which improved the accuracy of the sunspot number records and thus irradiance reconstructions \citep{kopp_impact_2016}.

More critical for historical reconstructions is, however, the lack of accurate information on the evolution of faculae and the network. 
This is particularly unfortunate, as changes in the network are believed to be the main source of the secular irradiance variability \citep{solanki_secular_2002}, which is the prime interest of climate studies.
Various models take different roundabout approaches to 
estimate
the possible changes in the facular component in the past.
Thus estimates of the change in TSI since the Maunder minimum reported in the literature diverge by about an order of magnitude (\citealp[see e.g.,][]{solanki_solar_2013-1}; cf. \citealp{yeo_dimmest_2020}).

Historical solar observations in the Ca II K
line have the unique potential of providing information on the evolution of faculae (or plage as they are called when seen in the chromosphere). 
Such observations started at the end of the 19th century and multiple observatories around the globe have initiated related programs.
The established link between the Ca~II~K brightness and photospheric 
magnetic field strength in bright magnetic regions \citep[e.g.][]{babcock_suns_1955,skumanich_sun_1984,schrijver_relations_1989,loukitcheva_relationship_2009,kahil_brightness_2017,kahil_intensity_2019,chatzistergos_recovering_2019} make these data an excellent proxy of solar magnetic activity.

Ca~II~K observations have, indeed, been used for irradiance modelling in the past.
Table \ref{tab:irradiancereconstructions} lists published models that employed Ca~II~K data for irradiance reconstructions.
Most of these models employed disc-integrated measurements of plage areas derived from analysis of a single modern archive of Ca~II~K observations taken with CCD devices and were limited to the last several decades.
Typically such models combine the plage area series with a sunspot record (number or areas) through a linear regression to match a record of irradiance measurements (see the first block of 
Table \ref{tab:irradiancereconstructions}). 
Another group of models followed the approach first used by \cite{johannesson_reproduction_1995} and \cite{chapman_variations_1996}, who considered the disc-integrated intensity in the Ca~II~K line as a facular proxy.  \cite{johannesson_reproduction_1995} used only Ca~II~K data, while \cite{chapman_variations_1996} used a disc-integrated intensity time-series produced from observations in the red (672.3 nm) continuum as a sunspot proxy in addition to the Ca~II~K data (see second block of Table \ref{tab:irradiancereconstructions}).
The more advanced semi-empirical models (last block of Table~\ref{tab:irradiancereconstructions}) employed the modern CCD observations to derive the locations of various magnetic regions on the surface. They then used the appropriate semi-empirical solar model atmospheres for these various features to assign the corresponding intensity spectrum to each pixel of the image. The brightness of all pixels is then summed up to give the irradiance.
A similar approach was applied by \cite{morrill_calculating_2005} and \cite{morrill_estimating_2011} to reconstruct the irradiance around the Mg II line (276--288 nm) but they used empirical center-to-limb (CLV, hereafter) functions derived from HRTS-9 \citep[High Resolution Telescope and Spectrograph;][]{brueckner_observations_1983} rocket observations instead of semi-empirical model atmospheres. Thus, this approach could only be used at the wavelengths where the CLVs were obtained.

To our knowledge, only three published TSI reconstructions are based on historical Ca~II~K photographic observations, all from Mt Wilson, and extend beyond the direct TSI measurements, going as far back as 1915 \citep{foukal_comparison_2002,foukal_new_2012,ambelu_estimation_2011}.  
They used a regression model with disc-integrated plage area series or Ca~II~K index as facular proxy. 
The series employed were derived from raw digitized images of the historical photographic observations without accounting for the problems plaguing such data \citep[see][]{chatzistergos_analysis_2018,chatzistergos_analysis_2019}.
Therefore no meaningful conclusion can be drawn about the secular irradiance variability from this analysis.

To fill this gap and ultimately reconstruct solar irradiance from direct facular observations over a period longer than several decades, we have started a comprehensive action on collecting and processing the available historical Ca~II~K data to bring them into the form needed to achieve more reliable irradiance reconstructions. 
\citet{chatzistergos_analysis_2018} developed a new method to accurately process historical and modern Ca~II~K observations. 
By applying this method on synthetic data, \citet{chatzistergos_analysis_2018,chatzistergos_analysis_2019} showed that it performs significantly better than other published methods.
In subsequent studies we have collected and processed numerous archives of Ca~II~K observations covering the period 1892--2019 \citep{chatzistergos_exploiting_2016,chatzistergos_ca_2018,chatzistergos_analysis_2018,chatzistergos_historical_2019,chatzistergos_analysis_2019,chatzistergos_delving_2019,chatzistergos_historical_2020,chatzistergos_analysis_2020}.
Furthermore, using a large sample of high-quality near co-temporal modern observations in Ca~II~K and magnetograms,
\cite{chatzistergos_recovering_2019} reassessed the relation between Ca~II~K brightness and magnetic field strength. 
This allows a reconstruction of maps of the unsigned magnetic field strength from Ca~II~K observations with the accuracy required for irradiance reconstructions.

Here we take the next step and reconstruct TSI variations over the period 1978--2019 with the Spectral And Total Irradiance Reconstructions model \citep[SATIRE, ][]{fligge_modelling_2000-1,krivova_reconstruction_2003,yeo_reconstruction_2014} adapted to exploit modern and historical Ca~II~K observations. 
The model and the data are described in Sects.~\ref{sec:satire} and \ref{sec:data}, respectively.
In Sect. \ref{sec:results} we present the reconstruction from modern data, while in Sect. \ref{sec:results_historical} we test the effect of some potential problems typical of historical data (such as variations in bandwidth, central wavelength, seeing as well as instrumental changes) on the quality of the reconstructions.
Finally we summarise our results and draw conclusions in Sect. \ref{sec:conclusions}.

\section{Model}
\label{sec:satire}

To calculate solar irradiance variations, we employ the SATIRE model
\citep[
][]{fligge_modelling_2000-1,krivova_reconstruction_2003,yeo_reconstruction_2014,yeo_solar_2017,dasi-espuig_reconstruction_2016,tagirov_readdressing_2019}.
SATIRE is a semi-empirical model.
It takes the distribution of the solar surface magnetic features (sunspot umbrae and penumbrae, as well as faculae and the network) from solar observations.
The brightness of these features is computed from the appropriate solar model atmospheres with the radiative transfer codes \citep{unruh_spectral_1999,norris_spectral_2017,tagirov_readdressing_2019}. 
The model 
replicates more than 90\% of the TSI variability measured from 1978 to the present \citep{yeo_reconstruction_2014,yeo_solar_2017}.

There are different versions of the SATIRE model, depending on the input data and consequently the period covered by the reconstruction \citep{krivova_towards_2011}. The ones of relevance for this paper are:\\[2pt]
-- SATIRE-S (for Satellite era; \citealp{krivova_reconstruction_2003,krivova_reconstruction_2009,wenzler_reconstruction_2006,ball_reconstruction_2012,yeo_reconstruction_2014}) and the most recent SATIRE-3D \citep{yeo_solar_2017} use full-disc longitudinal magnetograms to derive facular coverage as a function of time and white-light observations for sunspots. Such data are available for the last four decades.\\
-- SATIRE-T and -T2 (for Telescopic era) use sunspot observations (areas, sunspot number, group sunspot number) to infer the coverage by both sunspots and faculae back to 1609 \citep{krivova_reconstruction_2007,krivova_reconstruction_2010,dasi-espuig_modelling_2014,dasi-espuig_reconstruction_2016,wu_solar_2018-2}.\\

We base our model on the SATIRE-S version  \citep[see][]{fligge_modelling_2000-1,krivova_reconstruction_2003,yeo_reconstruction_2014}.
In SATIRE-S,
irradiance variations are computed as the sum over the entire solar disc of the intensity spectra 
of each image pixel.
Using the full-disc magnetograms and continuum images, each pixel is assigned in one of the four atmospheric components: sunspot umbra, penumbra, facula (including the network) and the quiet Sun (i.e. the surface free of any magnetic signal above the magnetogram noise threshold).
We use the same 
intensity spectra as in all standard versions of SATIRE-S, and SATIRE-T/T2 
\citep{wenzler_reconstruction_2006,ball_reconstruction_2012,yeo_reconstruction_2014,dasi-espuig_reconstruction_2016}.
(Different facular spectra  were employed by \citealt{tagirov_readdressing_2019}, who used facular spectra computed in non-LTE approximation to test non-LTE effects on the reconstruction of the UV irradiance and by \citealt{yeo_solar_2017} in the SATIRE-3D version, who used intensities computed from 3D MHD solar atmospheres.) 
The spectra we employ were computed with the ATLAS9 
code \citep{kurucz_atlas_1970} by \citet{unruh_spectral_1999} from semi-empirical model atmospheres.
The quiet Sun is represented by the \citet{kurucz_new_1993,kurucz_atlas12_2005} model at T=5777 K, sunspot umbrae and penumbrae by the models with T=4500 K and T=5450 K, respectively and faculae and network by the FALP model \citep{fontenla_calculation_1999} modified by \citet{unruh_spectral_1999}.

Instead of direct solar magnetograms, which are only available for the last several decades and cannot be used for longer irradiance reconstructions, we use Ca~II~K images to derive the information on the distribution of the facular (plage) regions 
on the visible solar disc.
For this, Ca~II~K observations have been converted into unsigned longitudinal magnetograms (see Sect.~\ref{sec:data}), which provide information on the location on the disc and the magnetic field density in facular and network regions. 
From the latter, we derive the facular filling factors (i.e. the fraction of a given pixel covered by a facular region) as done in SATIRE-S \citep[see][]{yeo_reconstruction_2014}.
This means that the filling factor of a given pixel, $a_{ij}^\mathrm{f}$, is proportional to the (radial) magnetic field density in that pixel $|B_{r,ij}|$,
until the saturation limit, $B_{\mathrm{sat}}$ is reached:
\begin{align*}
	a_{ij}^\mathrm{f} &=
	\begin{cases}
		|B_{r,ij}|/B_{\mathrm{sat}}        & \text{if } |B_{r,ij}| < B_{\mathrm{sat}}
		\\
				1        & \text{if } |B_{r,ij}| \ge B_{\mathrm{sat}}.
	\end{cases}
\end{align*}

$B_{\mathrm{sat}}$ is the free parameter of the model.
It accounts for the uncertainty in the exact relationship between the magnetic flux measured in the magnetograms \citep[which depends on their spatial resolution, e.g.][]{krivova_effect_2004} and the true brightness of the corresponding magnetic region on the Sun \citep[for further details, see][]{fligge_modelling_2000-1,yeo_reconstruction_2014}.
(The SATIRE-3D version of the model uses 3D MHD solar atmosphere models to obtain this relationship for SDO/HMI magnetograms directly, \citealt{yeo_solar_2017}, but this approach is not possible when using Ca~II~K data, as the MHD simulations used for this purpose do not yet cover the heights at which the Ca~II~K line core is formed with sufficient accuracy.)
To fix the value of $B_{\mathrm{sat}}$, we search for the value that returns the TSI reconstruction with the lowest RMS difference to a reference record of TSI measurements (see Sect.~\ref{sec:pubseries}).

Sunspots cannot be adequately identified in the Ca~II~K data for the objective of our study.
Therefore, to derive the filling factors of umbrae and penumbrae, $a^{u,p}_{ij}$,
we use the sunspot area composite series by \cite{mandal_sunspot_2020}. 
This record of sunspot group observations had been compiled using historical data from various observatories covering the period 1874--2019, that is the whole period, over which Ca~II~K images are available.
This database provides areas and $\mu$ positions of individual sunspot groups, but not for umbra and penumbra separately.
We use a constant ratio of 0.25 for the umbra to penumbra area to separate between them \citep{solanki_sunspots_2003,wenzler_can_2005}. 
The filling factors of umbra and penumbra have the value of 1 in pixels belonging to these regions and 0 otherwise.

However, sunspots can be  present in Ca~II~K observations too. Their appearance depends
on the observational conditions and the bandwidth used for the observation \citep{chatzistergos_analysis_2019,chatzistergos_analysis_2020}.
Mostly, their contrasts are either negative or very close to zero, but some parts might still appear bright in Ca~II~K.
Since we get sunspot and facular information from different sources, it can happen that these bright sunspot patches are counted by us as faculae.
Their overall fractional contribution is, however, minimal.
In the following, unless otherwise stated, we 
assume that this effect can be neglected (cf. Sect.~\ref{sec:sunspotcontribution}).

\section{Data}
\label{sec:data}
In this work we employ observed full-disc Ca~II~K and red-continuum images as well as published time-series of TSI plus sunspot area series.

\subsection{Full-disc observations}
\label{sec:data-fulldisc}

We use full-disc Ca~II~K observations from the Baikal (Ba), Big Bear (BB), Brussels (Br), Calern (CL), Kanzelh\"ohe (Ka), 
Mees (MS), Meudon (taken with a spectroheliograph, MD1), Mt Wilson (MW), 
Rome (RP, taken with the Precision Solar Photometric Telescope, PSPT), 
San Fernando (SF, taken with the Cartesian Full-Disk telescope 2, CFDT2), and Teide (Te, taken with the Chromospheric Telescope, Chrotel) observatories. 
We also use observations taken in the line-wings (hereafter referred to as off-band) from Meudon and Mauna Loa (taken with the PSPT). In particular, we use observations from the Mauna Loa solar observatory centred at 3936.3 \AA~(MLW) as well as Meudon observations centred at 
3932.3 \AA~(MDW, in the blue wing of the line).
Table \ref{tab:observatories} lists the main characteristics of the analysed data.

The 13 archives considered in this study describe the main characteristics of the various available Ca~II~K series.
The full-disc Ca~II~K observations have been taken with a spectroheliograph (MD1, MDW, and MW) or an interference filter (Ba, BB, Br, CL, Ka, MLW, MS, RP, SF, and Te).
There are data taken exclusively with a CCD camera (Ba, BB, Br, CL, Ka, MDW\footnote{MDW has also data stored on photographic plates, however here we restricted our analysis to the CCD-based data.}, MLW, MS, RP, SF, and Te) or stored on photographic plates (MW), while MD1 includes both. 
The analysed archives cover the full range of employed bandwidths at the various sites, with SF and MD1 using the broadest and narrowest ones, respectively.
We also test CCD data with small (e.g., RP or Te) and large (e.g., MS or BB) pixel scale. 
We note that for MD1 we analysed newly digitised data over 1982 and 1997--2002. This was because many photographic plates  from the former digitisation over that period were found to exhibit saturated regions \citep{chatzistergos_analysis_2019}, which limit the applicability of those data for TSI reconstructions.

The Ca~II~K data used here encompass 63,691 images covering 13,372 days over the period 1978--2019. 
Since the main aim of this work is to develop and test a method for the irradiance reconstruction from Ca~II~K data, here we restrict the analysed period to that of the direct TSI measurements, i.e. since 1978.
The analysis by \citet{chatzistergos_analysis_2020} of the same archives over the considered period 
includes 73,819 images. 
However, for 414 days (mainly over 2018--2019),
the series by \citet[][see Section \ref{sec:satire}]{mandal_sunspot_2020} does not have information on the sunspot areas and hence we cannot use the Ca~II~K observations on these days for the irradiance reconstructions. 
The archives in general include a varying number of observations per day, 
with most observatories aiming at acquiring roughly 1--3 Ca~II~K images per day.
However, some observatories also carry flare patrol campaigns in H$\alpha$ line, which require a greater amount of data taken with shorter cadence, typically of a few seconds to minutes, and complement those with similarly high-cadence Ca~II~K observations.
That is the case for the Ba, Br, CL, Ka, and Te archives.
We analysed all available observations from Br. 
For Ba, CL, Ka, Te we either analysed the ``best'' observation per day as identified by the observers, e.g. Ba, or randomly selected 2--3 images devoid of artefacts due to clouds \citep[for more information, see][]{chatzistergos_analysis_2020}.
We did one exception for the first week of June 2014, for which we analysed all available observations from all datasets used here in order to test the influence of seeing on our results (see Sect.~\ref{sec:seeing}).
Over the analysed period, the high-cadence archives provide up to 18 and 868 images per day (for Ba and Br, respectively). 

We process  the  Ca~II~K images  with the methods described by \cite{chatzistergos_analysis_2018,chatzistergos_analysis_2019,chatzistergos_analysis_2020} to perform the photometric calibration (only for the historical data) and compensate for the limb-darkening and any large-scale intensity pattern, thus creating contrast images. 
All images have been processed in a consistent manner. 

We convert the Ca~II~K observations from all archives to unsigned longitudinal magnetograms with the relation determined by \cite{chatzistergos_recovering_2019}.
We note that this process returns SDO/HMI-like magnetograms because the relation was determined for SDO/HMI \citep[Helioseismic
and Magnetic Imager aboard the Solar Dynamics Observatory; ][]{scherrer_helioseismic_2012,schou_design_2012,pesnell_solar_2012} magnetograms. Therefore, some differences are expected between our reconstructed magnetograms and those obtained from sources other than SDO/HMI.
Since this relation was derived for RP data, to apply it on observations from other archives we normalised the standard deviation of the quiet Sun regions to the average value found for RP data.
In Figure \ref{fig:histograms} we compare histograms of the unsigned radial magnetic field density $|B_{r}|$ on 11 August 2001 (top panel), 16 July 2009 (middle panel), and 10 May 2012 (bottom panel) for the magnetograms reconstructed from RP, MD1, and SF datasets as well as the SDO/HMI magnetogram (shown only in the bottom panel as HMI data are not available prior to 2010). All images were first resized to the same dimensions. We also show the histograms for the magnetograms reconstructed from Ca~II~K images without their normalisation to the RP quiet Sun contrast. The  normalisation improves considerably the agreement with the RP data.
The histogram for the SDO/HMI magnetogram on 10 May 2012 is very close to the histograms for the reconstructed magnetograms after normalising them to RP data. 
We note that the $3-\sigma$ noise level of the 315-second SDO/HMI magnetograms is $\sim$20 G \citep{yeo_intensity_2013,yeo_reconstruction_2014}, which explains the difference at lower $|B_{r}|$ values.

However, we also notice some mismatch, mostly at high $|B_{r}|$ values, potentially hinting at a non-linear relation of the normalisation factors between archives. 
We do not expect the mismatch over high $|B_{r}|$ values to be important for our analysis here since the free parameter of the model $B_{\mathrm{sat}}$ would account for this.
But a forthcoming study will address this issue in more detail.

To test the effect of sunspot identification (Sect.~\ref{sec:sunspotcontribution}), we also analyse full-disc RP observations taken in the red continuum at 607.1 nm with a bandwidth of 0.5 nm covering the period 1997--2019. These images were processed in the same way as 
Ca~II~K images to compensate for the limb-darkening and large-scale intensity patterns affecting the observations \citep[][]{chatzistergos_modelling_2020}.

\subsection{TSI series} 
\label{sec:pubseries}

To constrain the free parameter of the model and to test the accuracy of our reconstructions, we also use three records of direct TSI measurements from individual instruments, four TSI composites of measurements, and 7 irradiance series produced with models.

In particular, we use the TSI measurements from the Total Irradiance Monitor (TIM) on-board the SOlar Radiation and Climate
Experiment\footnote{Available at the LISIRD archive at \url{lasp.colorado.edu/lisird/}} \citep[SORCE, version 19, February 2020 both for daily averages and 50-s cadence data;][]{kopp_total_2005}, TSI Continuity Transfer Experiment$^\thefootnote$ \citep[TCTE/TIM, version 4 for 50-s cadence data;][]{kopp_tcte_2013}, and Variability of solar IRadiance and Gravity
Oscillations (VIRGO) experiment on-board the SOlar and
Heliospheric Observatory\footnote{Available at \url{https://www.pmodwrc.ch}} \citep[SOHO, version 6.5, May 2018 for daily averages and version created on 09/06/2016 for 60-s cadence data series;][]{frohlich_-flight_1997} instruments. We note that the daily values were used to constrain the free parameter, while the high-cadence data were used only to compare to our reconstruction. 

The composite TSI series used are the ACRIM\footnote{Available at \url{http://www.acrim.com}} \citep[Active Cavity Radiometer Irradiance Monitor, which is the instrument taken as the reference by][version 30/11/2013]{willson_total_1997,willson_composite_2003}, \addtocounter{footnote}{-1} PMOD$^\thefootnote$  \addtocounter{footnote}{1} \citep[named after Physikalisch-Meteorologisches Observatorium Davos; version 42.65 15/07/2018;][]{frohlich_solar_2006}, RMIB\footnote{Available at \url{ftp://gerb.oma.be/steven/RMIB_TSI_composite/}} \citep[named after Royal Meteorological Institute of Belgium, in French called IRMB, version downloaded on 17/08/2020;][]{dewitte_total_2004,dewitte_total_2016}, and the one by \cite[][GEA18, hereafter]{gueymard_reevaluation_2018}. 
The ACRIM, PMOD, and RMIB composites are created with a daisy-chain process of using the measurements of one instrument as the reference to calibrate the counts of a  second one, which then would act as the reference for another instrument. 
A value provided on a given day is an average over all measurements on that day by a single instrument.
We note that the ACRIM TSI composite series has not been updated since 2013,
but it is used here for completeness.

For comparison we also use 7 irradiance series produced with models. These are the TSI reconstructions with the SATIRE-S\footnote{Available at \url{http://www2.mps.mpg.de/projects/sun-climate/data.html}} \citep[covering 23/08/1974--21/06/2019]{yeo_reconstruction_2014}, SATIRE-T$^\thefootnote$ \citep[covering 02/07/1643--31/05/2017]{wu_solar_2018-2}, SATIRE-T2$^\thefootnote$ \citep[covering 03/01/1700--07/11/2008]{dasi-espuig_reconstruction_2016},
\addtocounter{footnote}{-4} the NRLTSI$^\thefootnote$ \citep[Naval Research Laboratory TSI, covering 01/01/1882--31/12/2020;][]{coddington_solar_2016},
\addtocounter{footnote}{4} the EMPIRE$^\thefootnote$ \citep[EMPirical
Irradiance REconstruction, covering 14/02/1947--31/05/2017;][]{yeo_empire_2017}, \addtocounter{footnote}{-4}  NN-SIM$^\thefootnote$ \citep[Neural Network for Solar Irradiance Modeling, covering 08/11/1978--31/03/2019;][]{mauceri_neural_2019}, \addtocounter{footnote}{4} and the \citet[][PSUM, hereafter, for Photometric SUMs, covering 16/05/1996--12/06/2020]{chatzistergos_modelling_2020}$^\thefootnote$  model.
The methodology of SATIRE-S, SATIRE-T, and SATIRE-T2 
is described in detail in Section \ref{sec:satire}.
The remaining models apply linear regressions between solar activity proxies and measurements of solar irradiance.
NRLTSI and EMPIRE both use the Mg II index for the facular contribution for the period discussed in this work, but differ in that NRLTSI applied ordinary least squares regression, while EMPIRE used orthogonal distance regression.
NN-SIM reconstructed irradiance using artificial neural
networks fed with 6 solar proxies: Mg II index, Lyman $\alpha$ irradiance at 121.6 nm, photometric sunspot index, and radio flux at 10.7, 15, and 30 cm.
The series by \cite{chatzistergos_modelling_2020} used the photometric sum indices computed from RP Ca~II~K and blue continuum observations for the regression, following \cite{chapman_modeling_2013}.

\section{Results for RP observations} 

\label{sec:results}

In this section, we present and discuss the TSI reconstruction from the RP data, which is the longest and most consistent modern archive of high-quality Ca~II~K observations.
Results for other archives are presented in Sect.~\ref{sec:results_historical}

\subsection{
Irradiance reconstruction and the effect of the reference TSI series
}
\label{sec:results_pspt}

Figure \ref{fig:pspt} shows the TSI reconstructed from the RP images with the PMOD TSI series used as the reference to set $B_{\mathrm{sat}}$. 
The model and the PMOD record are in good agreement with each other.
Pearson's linear correlation coefficient, $R$, and the RMS difference
over the whole period
are 0.91 and 0.21 Wm$^{-2}$, respectively.
For comparison, the correlation coefficient and the RMS difference between the PMOD and SATIRE-S 
over the same period
are 0.98 and 0.16 Wm$^{-2}$, respectively.
To better illustrate the agreement between the two models and the PMOD record on shorter timescales,
in Fig.~\ref{fig:psptdaily} we also show the TSI changes 
over two 2-month periods close to the activity maximum (June--July 2001) and minimum (June--July 2009).
While the Ca~II~K-based reconstruction appears slightly noisier than SATIRE-S relying on space-based magnetograms, it still does an excellent job.

It has already been shown earlier \citep{wenzler_reconstructed_2009,krivova_acrim-gap_2009,yeo_reconstruction_2014,chatzistergos_reconstructing_2021-1} 
that the choice of the TSI reference record does not significantly affect the value of the free parameter, $B_{\mathrm{sat}}$, and thus the final reconstructions.
Here, we nevertheless test this again.
In particular, we use four TSI composites (PMOD, ACRIM, RMIB, and GEA18), two individual instrumental records (SOHO/VIRGO and SORCE/TIM), as well as seven alternative TSI models (SATIRE-S, SATIRE-T, SATIRE-T2, EMPIRE, NN-SIM, NRLTSI and PSUM) as references (see Sect.~\ref{sec:pubseries} for the description of these data).
Table~\ref{tab:tsibsat}
lists the resulting $B_{\mathrm{sat}}$ values together with the corresponding $\mathrm{R}$ and RMS for these various reconstructions.
In particular, the best agreement with PMOD is reached with  $B_{\mathrm{sat}}=201$\,G.
This value is roughly in the middle of the entire range of 188--212\,G  obtained 
for all reference series of direct TSI measurements considered here. 
The RMS differences between the reconstruction with $B_{\mathrm{sat}}=201$ G and those with the two extreme cases of  $B_{\mathrm{sat}}$ are 0.043 and 0.010 Wm$^{-2}$, respectively.
Somewhat higher values of $B_{\mathrm{sat}}$, 238 and 263\,G, are obtained when using the SATIRE-T and SATIRE-T2 models, respectively. These somewhat higher deviations are attributed to the concept of those two models. They rely on sunspot observations alone and are less accurate on time-scales of days to a few years \citep{krivova_acrim-gap_2009}.

Nevertheless, employment of all records listed in Table~\ref{tab:tsibsat} as the reference results in very similar TSI reconstructions (see Appendix \ref{sec:appendixcomparison}).
This is also demonstrated in Fig.~\ref{fig:pspt} and \ref{fig:psptdaily} by the shaded area which marks the entire range of the reconstructed TSI values.
The value of $B_{\mathrm{sat}}$ affects mainly the amplitude of the cycle variation, with higher values of $B_{\mathrm{sat}}$ returning weaker cycle amplitudes.
The long-term trend is, however, barely affected by the selection of $B_{\mathrm{sat}}$ within the range of values derived with the various reference series \citep[see also][]{chatzistergos_reconstructing_2021-1}. 
The correlation coefficient, $R$, is $\sim$0.9 or higher, while the RMS stays below 0.25~Wm$^{-2}$ in all cases.
The only exception for the direct TSI series is the ACRIM composite, with which the agreement is clearly poorer (R=0.81 and RMS=0.35).

In summary, the TSI reconstructed from the RP data is in excellent agreement with the published TSI series (except ACRIM), and
the result is essentially independent of the series employed as the reference record to determine the free parameter.

\subsection{Contribution of sunspots to facular filling factors}
\label{sec:sunspotcontribution}

As mentioned in Sect. \ref{sec:satire}, sunspots can also partly appear as bright regions in Ca~II~K observations. 
Since we derive sunspot and facular filling factors from different sources, we might count some sunspot regions as faculae.
To check the potential effect of the miscounted sunspots on the model outcome, we performed and compared the following three runs: 
\begin{enumerate}
	\item The sunspot series by \cite{mandal_sunspot_2020},  which gives the $\mu$ values of the sunspot groups but not their exact locations, was used to compute the filling factors of sunspots (this is the approach used in Sec. \ref{sec:results_pspt} and \ref{sec:results_historical}).
	\item The series by \cite{mandal_sunspot_2020} was used to compute the filling factors of sunspots, and the latter were then subtracted from the facular filling factors for the same $\mu$ positions. 
	\item The RP red-continuum observations were used to identify sunspots and derive their filling factors and positions. These exact locations of sunspots were then used to block these regions in the Ca~II~K observations before computing facular filling factors.
\end{enumerate}
The first two cases make the extreme assumptions that sunspots either do not appear as brightenings in Ca~II~K data at all (case 1) or that they are always seen (case 2). Case 3 is the most accurate case, as we take exact positions and areas of spots into account.

Figure \ref{fig:psptdifsunspotdif} shows the TSI reconstructed from RP 
with these different approaches to account for the sunspots, while scatter plots of these reconstructions versus the PMOD composite are shown in Fig.~\ref{fig:pspt11scatter}.
All reconstructions are remarkably similar, and 
the effect of the 3 different approaches to account for sunspots on the resulting TSI is clearly minor. In particular, while the RMS differences ($R$) between the reconstructed series and PMOD TSI series is 0.20 (0.92) when using the RP red-continuum observations, the values are 0.21 (0.92) in both cases when we use the series by \cite{mandal_sunspot_2020}. 
These results remain qualitatively unchanged even if Ca~II~K datasets taken with different bandwidths are considered (see Appendix \ref{sec:bandwidth_sunspot}).

These results suggest that the simplest approach of using the sunspot area series and the facular filling factors derived from Ca~II~K observations
without correcting for the visibility of spots in the Ca~II~K data is entirely adequate to our purpose.
This is important for historical irradiance reconstructions, because this approach can be used even with Ca~II~K data for which we do not have co-temporal continuum observations.

\section{Towards TSI reconstructions from historical Ca~II~K archives}
\label{sec:results_historical}

In the previous section we have shown that on modern high-quality Ca~II~K observations, our model works very well,
reproducing $\gtrsim 85\%$ of the measured TSI variations.
This brings us one step closer to a reconstruction of the historical solar irradiance variability using Ca~II~K images to obtain facular coverage.
Unfortunately, historical Ca~II~K observations have lower quality than the RP data and suffer from various issues.
While our method \citep{chatzistergos_analysis_2017,chatzistergos_analysis_2018,chatzistergos_analysis_2019,chatzistergos_analysis_2020} allows for resolving many of these issues, the quality of such data remains poorer than that of RP images.
Therefore, additional work will be needed to allow a reliable irradiance reconstruction back to the late 19th century.
Here, we discuss such issues and their potential effect on the reconstructions.
We start by considering the effect of the pixel scale of observations on the results (Sect.~\ref{sec:results_pixelscale}).
Another critical issue is that observations have been taken with different bandwidths or even off-band (Sect.~\ref{sec:results_ccddata}).
We also consider the effects of instrumental changes (Sect.~\ref{sec:results_discont}) and of varying or poor seeing (Sect.~\ref{sec:seeing}) on the reconstructions.
Finally, we show two examples of reconstructions from photographic archives (Sect.~\ref{sec:results_photo}) and discuss the possibility of using such a reconstructed series, in turn, as a reference for reconstructions from other archives (Sect.~\ref{sec:sensitivity_reference}).

\subsection{Effect of the pixel scale of the observations}
\label{sec:results_pixelscale}

Observations from different archives differ considerably in their pixel scale, ranging from 1 to 5.5"/pixel for the archives analysed here.
To test whether the pixel scale affects our TSI reconstructions, we used the RP Ca~II~K data (with the original pixel scale of ~2"/pixel)
and performed eight TSI reconstructions by resizing the observations such that the pixel scale ranged between 2.0 and 9.6"/pixel.
We used the RP data,
because of their quality, consistency and sufficiently long period of observations.

The reconstructions using the resized RP data are compared to that from the original RP data and to the PMOD TSI composite  in Table \ref{tab:tsibsat_resized}.
The value of $B_{\mathrm{sat}}$ grows somewhat with the increasing pixel scale from 201\,G for the original-size data up to 248\,G for the extreme case of 9.7"/pixel.
The differences in the quality of the reconstructions between the various cases is essentially negligible for pixel scales up to 4.9"/pixel, while the quality is still rather good for a pixel scale of 9.7"/pixel.
In particular, while the RMS difference between the resized and original reconstructions grows with the pixel scale, it remains low,
with merely 0.081Wm$^{-2}$ for the pixel scale of 9.6"/pixel.
Compared to PMOD, the RMS difference again changes very little for the resized data, reaching 0.223Wm$^{-2}$  for the pixel scale of 9.6"/pixel (0.214Wm$^{-2}$ for the reconstruction with the original RP data).
This demonstrates that 
the pixel scale of the observations affects our results only weakly and suggests that data from archives with a large pixel scale, such as those from MS and BB, can also be used for irradiance reconstructions. 

This is one of the big advantages of using Ca~II~K data compared to magnetograms. Due to the cancellation of opposite polarities, the magnetogram signal, and hence the amount of magnetic flux that is detected, drops rapidly with decreasing spatial resolution. Ca~II~K plage and the absolute magnetograms, including the absolute flux that are obtained from them, are far less sensitive to spatial resolution.

\subsection{Effect of the band-pass of the observations}
\label{sec:results_ccddata}

We now consider CCD data taken with filters with different bandpasses (see Table \ref{tab:observatories}). 
Whereas Rome (RP) data were taken with a 2.5\AA\ wide filter,
the data from Baikal (Ba), Mees (MS), and Teide (Te) were recorded with narrower filters in the range 0.3--1.2\AA.
Broader filters with a width between 2.7 and 9\AA\ were used at
Brussels (Br), Calern (CL) and San Fernando (SF).  
Figure \ref{fig:tsidaily_modern} shows the reconstructed TSI from the above six Ca~II~K archives 
along with the PMOD TSI composite, while
Table~\ref{tab:tsibsat} lists the resulting $B_{\mathrm{sat}}$ values together with the corresponding $R$ and RMS for the various reconstructions.
We focus here on the case when PMOD is used as the reference.
But for completeness, Table~\ref{tab:tsibsat} also lists the cases when alternative reference records are used (as discussed in Sect.~\ref{sec:results_pspt} for the example of RP reconstructions).

While the most accurate reconstruction is clearly achieved from the RP data, the agreement with the PMOD composite is good for all considered Ca~II~K archives.
$R$ is above 0.78 and the
RMS difference is less than 0.32 Wm$^{-2}$ 
for all considered datasets.
Overall, the reconstructions from data taken with a broad or narrow bandwidth and their quality are all comparable.
Some discrepancy between the TSI values in the reconstruction with MS data 
and the PMOD values in the early 1990's is due to scarcity of the data during that period, such that a meaningful comparison is difficult.
Furthermore, the spectral bandwidth of the MS data was, in fact, not constant over the time, varying between 0.3 and 1.2\,\AA.
Except MS, the highest RMS differences resulted from CL, SF and Te data.
SF and Te are the two extreme cases with respect to the employed bandwidth among the CCD-archives tested here, SF using the broadest and Te the narrowest bandwidth.
The best estimate of $B_{\mathrm{sat}}$ generally increases with decreasing bandwidth employed for the Ca~II~K observations.
In particular, the adopted $B_{\mathrm{sat}}$ value for RP data (201 G) is lower than that for Te data (352 G), which is consistent with the Te data having a narrower bandwidth, 
than the RP data.

To analyse the effect of the off-band observations on the irradiance reconstructions, we have considered the data from Meudon (MDW) and Mauna Loa (MLW). 
 Figure~\ref{fig:tsidaily_offbandmodern} and Table \ref{tab:tsibsat} present TSI reconstructions from these archives.
These reconstructions are also in good 
agreement with the PMOD TSI series.
But we again notice that the adopted $B_{\mathrm{sat}}$ generally decreases with the offset of the central wavelength of the observations from the line core. 
This might be due to the non-linearity of the relation between archives with different bandpasses (see Fig.~\ref{fig:histograms} and Sect.~\ref{sec:data-fulldisc}), which will be addressed in a separate paper.

Overall, our preliminary analysis suggests that data taken with a narrower or broader passbands or offsets in the central wavelength can generally be used for historical reconstructions.
Also noteworthy is that the accuracy of the reconstruction with the MS data is similar to that of the other archives (keeping in mind the scarcity of the observations), highlighting the value of even the archives with large pixel-scale for this purpose  (see Sect.~\ref{sec:results_pixelscale}).

\subsection{Uncertainties due to instrumental discontinuities}
\label{sec:results_discont}

Another issue one has to deal with when using historical observations are changes in the instrumental set-ups with time.
As an example, in Fig.~\ref{fig:tsidaily_modernBBML} we show the TSI reconstructed from BB and Ka data, along with the PMOD TSI composite for comparison. 
Both archives have known issues affecting consistency of their quality \citep{chatzistergos_analysis_2020}.

The TSI series reconstructed from the BB data 
shows clear and abrupt changes, particularly evident in the difference between the PMOD TSI series and the reconstruction. 
These jumps coincide perfectly with the periods of instrumental changes (marked by vertical dashed lines in the figure).
Most prominent is the jump on 07 July 2000, when a CCD calibration started being applied to the data.
A possible exception is 10 September 1996 
when the BB filter was replaced. At this time we do not see any abrupt change in the values compared to those from the PMOD series, although the scatter seems to decrease after the change.

The reconstruction from Ka data clearly deviates from PMOD TSI over 2011--2012, and the linear correlation drops down to 0.32 over this period.
This coincides with the deterioration of the initial filter used at Ka, leading to a poorer quality of the images over this period. 
The filter was replaced on 24 November 2012. 
The results after the replacement of the filter at Ka are in excellent agreement with the PMOD TSI series.

These examples show the importance of understanding and knowing, as precisely as possible, any instrumental issues affecting the Ca~II~K datasets.
If not properly accounted for, they will result in artefacts in the computed TSI.
This also means that
 archives with such inconsistencies, which are not precisely accounted for, should be avoided when studying the long-term trend in solar irradiance.
 Furthermore, this points to the importance of using multiple archives whenever possible, which might help to identify inconsistency that would otherwise not be noticed.

\subsection{Uncertainty due to sampling and seeing}
\label{sec:seeing}

The TSI reconstructions presented in the previous sections were daily mean values over all individual images (from a given archive) within a day. In this section, we discuss  reconstructions from {\em individual} images from Ba, Br, CL, Ka, and Te datasets and compare them to  the SOHO/VIRGO record with an hourly cadence as well as to SOHO/VIRGO, SORCE/TIM, and TCTE/TIM series with an approximately 1-minute cadence.
Figure~\ref{fig:201406}a shows the reconstructed TSI over the course of the first week of June 2014.
Panels b, c and d are enlargements of panel a over 05 and 07 June 2014, as well as between 13:00 and 14:00 UTC on 07 June 2014, respectively. 
The variability within a given day is similar in the reconstructed TSI and
in the measurements done with the 1-minute cadence.
The main sources of TSI fluctuations on these timescales are 5-minute oscillations \citep{leighton_velocity_1962} and granulation \citep[see, e.g.,][]{shapiro_nature_2017},
which we do not expect to see in Ca~II~K data.

Generally, on time-scales of hours to a few days,  the computed variability mimics that in the SOHO/VIRGO series, as can be seen for Ka over 7 June 2014 or Te over 5 June 2014.
However, we sometimes also notice variations within a day with trends  different from those in SOHO/VIRGO, e.g. for Te over 6 June 2014 or 7 June 2014.
This is most probably the result of changing seeing conditions during the day at the Teide site.

For a more quantitative comparison, we
remove roughly the effect of the evolution of magnetic features from all the reconstructed TSI series in the following way.
We smooth  the SoHO/VIRGO data having the 1-minute cadence with a 30-minute window
and then subtract this smoothed series both from the original 1-minute record and from all reconstructions.
We then compute the standard deviations of these residual series within each day and list them in Table \ref{tab:tsibsatdaily}.     
For the considered first week of June 2014, the values for individual days and data sets lie in the range [0.03--0.18] Wm$^{-2}$.
In particular, the values for the measurements lie between 0.05 and 0.11, while those for the reconstructions between 0.03 and 0.18.
The values for the reconstructions are generally comparable to those for the measurements, although with a tendency to be slightly higher. 
They also vary more from day to day than the values from the measurements.
On 6th June, three of the four archives (Br, CL and Te) result in standard deviations more than twice higher than in the measurements (0.16--0.18 vs. 0.07).

We consider these slightly higher values of the standard deviation as a very rough estimate of the uncertainty in the reconstructed TSI due to seeing.
We stress, however, that this is not a robust measure  since it includes contributions from other artefacts plaguing the images too.

\subsection{Reconstructions using photographic Ca~II~K archives}
\label{sec:results_photo}

So far, we have considered CCD-based archives.
Earlier observations have, however, been stored on photographic plates and films.
Compared to CCD observations, analysis of photographic archives involves more intricacies.
Their main sources are
(1) the non-linear response of photographic material to sunlight exposure, and (2) inconsistencies within the datasets due to changes in the instrumentation and settings for the digitisation unit. 
Photographic archives have predominantly employed a spectroheliograph, which allows for greater versatility in the observing parameters and many observatories took advantage of this. This, however, resulted in many archives being collections of observations with potentially varying settings. Such changes have not always been noted down or this information has not been transferred to a digital form.
Furthermore, inconsistencies within the archives are potentially introduced due to changes and updates in the instrumentation used for the observations.
Such inconsistencies are rather common, given the long duration of the observing programs, which in a few cases, such as  MD1, lasted for more than a century.
It is worth mentioning that images from photographic archives also suffer from large-scale artefacts. These have, however, been efficiently compensated for by the image processing we had applied to the data

Next, as a fundamental step towards the application of our method to historical Ca~II~K observations, we now consider two examples of historical photographic archives, MD1 and MW.
We focus on the period after 1978, over which direct TSI measurements are available for comparison.
The reconstructed TSI from the MD1 and MW archives is shown in
Fig.~\ref{fig:tsidaily_historical}, and the
results of the quantitative comparison to the PMOD record are listed in Table \ref{tab:tsibsathistorical}.
For comparison, we also reiterate in the table the results for the reconstruction from RP CCD-based observations.

The RMS differences between the reconstructed TSI series and the PMOD record are generally higher for the historical photographic data than for the CCD-based ones. 
The reconstruction from MW data looks fairly good, with RMS differences to PMOD TSI series of 0.43.
Some deviation in the long-term trend is seen, which might be because the considered period (i.e. overlap with satellite TSI data) is quite short to reliably fix the value of the free parameter $B_\mathrm{sat}$.
Unfortunately, there are also known inconsistencies in the MW series before and after 1976, which
makes an extension of this reconstruction further back in time not straightforward. 
The reconstruction from the MD1 archive is generally of comparable quality to that from MW, although
the scatter in the residual is higher during activity maxima.
In contrast, the long-term trend is in good agreement with that in PMOD.
But the reconstruction exhibits some abrupt changes during known inconsistencies in the data, e.g
over 1980 when the digitisation of MD1 data changed, or in 2002 when a CCD camera started being used at MD1. 

In Table \ref{tab:tsibsathistorical} we also compare our reconstructions from the historical archives and the CCD-based RP archive to that from the SATIRE-S model, which uses the same method but is based on direct magnetograms  as well as those from the SATIRE-T and SATIRE-T2 models. 
 To allow for a direct comparison, we also
 list the RMS difference between the various models and PMOD on the same dates as for the respective Ca~II~K series. 
This comparison is restricted to the period prior to 07 November 2008, which is the period covered by all TSI reference series used here.
 Unsurprisingly, the reconstruction from the observed magnetograms (i.e. with SATIRE-S) performs better than those from the historic archives.
Nevertheless, even with the historic archives, the correlation coefficient with the directly measured TSI is over 0.75.
Compared to SATIRE-T/T2 our reconstructions from modern data fare better, while the quality of the reconstructions from historical data is comparable to that of the SATIRE-T/T2 reconstructions.

To summarise, our results show that photographic observations from historical Ca~II~K archives can be used for past irradiance reconstructions.
To allow for reliable reconstructions, however, it is really important to understand and address the various instrumental issues affecting such data, as well as to
use observations from multiple archives processed consistently and accurately.

\subsection{Sensitivity of reconstruction to reference TSI series}
\label{sec:sensitivity_reference}

Since most of the historical archives will not have a direct overlap with irradiance measurements, one will have to use intermediate archives to ``cross-calibrate'' the reconstructions.
Therefore, we now also test whether we can accurately reconstruct TSI when using one of our reconstructed TSI series as the reference.
For this, we use the TSI series reconstructed from MD1 Ca~II~K data, which is the longest series analysed here, as the reference.
We note that the MD1 reconstruction was produced by considering Ca~II~K data stored on photographic plates (before 2002) as well as those taken with a CCD camera (since 2002).
We remind that the MD1 reconstruction itself was obtained using PMOD TSI as the reference.
The results for six of the archives considered by us are shown in Fig. \ref{fig:tsidaily_meudonref}.
In the top panel of each block we show the reconstruction for a given archive using MD1 as the reference as well as the PMOD TSI for comparison. The corresponding bottom panels show the difference between the PMOD TSI series and our reconstruction.

Table \ref{tab:tsibsat} also lists the obtained $B_{\mathrm{sat}}$ value as well as the RMS difference and linear correlation coefficient with the MD1 TSI series.
The RMS differences and linear correlation coefficient are given as compared to both the MD1 and the PMOD TSI series.
The RMS differences between the series reconstructed with PMOD and MD1 are generally comparable.
This is important for employment of the historical Ca~II~K data for irradiance reconstructions, because it suggests that we can use some of our reconstructed series as a reference for Ca~II~K archives that do not directly overlap with actual measurements of TSI.

\section{Summary and conclusions}
\label{sec:conclusions}

Reconstructions of solar irradiance variations on timescales of decades to centuries are usually carried out using sunspot observations.
Such observations, however, provide only indirect information on the evolution of the bright magnetic features (faculae or plage, when observed in the photosphere or chromosphere, respectively).
Since the late 19th century, the Sun has also been regularly observed in the Ca~II~K line, whose brightenings directly reflect the presence of plage and network. Consequently, the brightness in this line on the solar disc is a good proxy of the photospheric magnetic field.
Due to numerous, partly severe, artefacts and problems affecting such data, the data cannot be used for irradiance or similar studies directly. \citet{chatzistergos_analysis_2018,chatzistergos_analysis_2019} have developed a novel, accurate method to account for most of these problems, which they then applied on images from 43 archives \citep{chatzistergos_analysis_2020}. 
Next, \cite{chatzistergos_recovering_2019} have analysed simultaneous high-quality, high-resolution RP Ca~II~K and SDO/HMI observations to derive the relationship between the Ca~II~K brightness and the magnetogram signal. 

Here we take the next steps towards exploring the potential of historic solar observations in the Ca~II~K line for irradiance reconstructions. 
Using full-disc Ca~II~K observations from various archives,
we have reconstructed the total solar irradiance, TSI, back to 1978.
To do so, we have converted Ca~II~K observations into unsigned magnetograms by using the relationship by \citet{chatzistergos_recovering_2019} and adapted 
the SATIRE-S model \citep{krivova_reconstruction_2003,yeo_reconstruction_2014}, which is normally fed by observed magnetograms, to use the unsigned magnetic maps resulting from the Ca~II~K images.

We have first extensively tested the model by using the modern, high-quality CCD-based Rome/PSPT observations and
showed that our model performs at levels comparable to the original SATIRE-S version.
We also showed that the reference TSI series, used to fix the free parameter of the model, $B_\mathrm{sat}$, does not affect significantly the reconstructed TSI variations.
For all independent direct TSI records considered by us as the reference, including TSI composites and measurements by individual instruments, the correlation coefficient with our model and the RMS difference lie in the range 0.90--0.91 and 0.17--0.23 W\,m$^{-2}$, respectively (Table~\ref{tab:tsibsat}).
The only exception is the ACRIM TSI composite, for which the agreement is significantly poorer, with a correlation coefficient of merely 0.81 and a RMS difference of 0.35~W\,m$^{-2}$.

For the period of time covered by Rome/PSPT observations, we can extract sunspot locations and areas from co-temporal red-continuum images.
For historical Ca~II~K archives, this is, unfortunately, not possible. Therefore, we have tested different methods for accounting for sunspots by using the record of historical sunspot observations by \citet{mandal_sunspot_2020}.
We find only a marginal improvement of the quality of the reconstructions when using co-temporal continuum observations to identify the location and area of sunspots instead of the series by \cite{mandal_sunspot_2020}. This is an important result, because it means we can combine historical sunspot observations with Ca~II~K image data to reconstruct past irradiance variability with hardly any loss of accuracy.

Next,
we employed 10 further CCD-based archives to test
the effects of the spatial pixel scale, spectral band-pass, changes in properties of the employed instruments, temporal sampling
and seeing conditions on the outcome of the model.
We estimated the uncertainty in the results due to seeing conditions to generally lie below 0.2 Wm$^{-2}$. 
We showed that we can obtain reasonably accurate results even with data with a large pixel scale.
This is another advantage of reconstructing solar irradiance with Ca II K observations instead of magnetograms.
The amount of detected magnetic flux in magnetograms is strongly diminished with decreasing spatial resolution and the resulting lower net flux due to sub-pixel regions of opposite polarities,
while the unsigned magnetograms reconstructed from Ca II K observations are less sensitive to the spatial resolution.

To construct the unsigned magnetograms, we applied to all archives the same relation, which was derived by \citet{chatzistergos_recovering_2019}
using RP data. 
Since some archives employed different bandpasses or used interference filters or spectroheliographs for the observations, this relationship might not be consistent with them.
We have used Ca~II~K data with bandwidths covering the entire range of available archives as input, as well as data with band-passes centred at the wings of the line (off-band), and showed that we could also obtain accurate irradiance reconstructions from such data by adapting the free parameter of the model, $B_{\mathrm{sat}}$. 
We have, however, noticed a dependence of the adopted value of the free parameter, $B_{\mathrm{sat}}$, on the bandwidth of the Ca~II~K observations, suggesting that the relationship we use to convert Ca~II~K data to magnetograms depends on the bandpass.
This might potentially be useful for irradiance reconstructions from such archives.
A detailed analysis of this effect
will be the focus of a forthcoming study.

Further, we present first tests of how well our model works when using historical photographic Ca~II~K archives. 
This is an important preparation for the use of the historical Ca~II~K archives to reconstruct TSI back to the end of the 19th century. We have used the images from the Mount Wilson (MW) and Meudon (MD1) archives to reconstruct the TSI over the period after 1978, when we can compare the outcome to direct TSI measurements.
We showed that inconsistencies within the archives can distort the results. However if the inconsistencies are identified, then such data can still be used to recover irradiance variations sufficiently accurately.
Thus, identifying and potentially accounting for such inconsistencies and discontinuities will be the main challenge for reconstructions of irradiance from historical archives.
Once such issues are better understood and taken into account for the available archives processed by \cite{chatzistergos_analysis_2020}, we will use them for the reconstruction of irradiance over the 20th century.

In this paper we focused on presenting results for TSI, however the model also produces the spectrally resolved solar irradiance (solar spectral irradiance, SSI) over 115 to 160,000 nm. Future work aiming at the reconstruction of the past irradiance variations from historical and modern Ca~II~K data using the method described in this paper will include both TSI and SSI. 

Finally, we used the TSI reconstruction from the MD1 archive (which were photographic prior to 2002 and CCD-based afterwards) 
as a reference for reconstructions from other archives.
We found that the quality of the reconstructions remained comparable to that when the PMOD TSI was used as a reference.
This is extremely important, considering that many historical datasets do not overlap with direct irradiance measurements and thus optimising the free parameter would not be possible otherwise.

In summary, we have shown that historical Ca~II~K data are a viable source of information on the facular and network contribution to solar irradiance variations. The availability of such data for over a century will allow reconstructing irradiance variations based on independent data sources for sunspots and faculae for a considerably longer period of time than was possible so far.

\begin{acknowledgements}
	The authors thank the observers at the Baikal, Big Bear, Brussels, Calern, Kanzelh\"ohe, Mauna Loa, Mees, Meudon, Mt Wilson, Rome, San Fernando, and Teide sites. We thank Isabelle Buale for all her efforts to digitise the Meudon photographic archive.
We also thank the anonymous referee for encouraging and constructive suggestions that helped us to improve the presentation of our work.
	T. C. acknowledges funding from the European Union's Horizon 2020 research and Innovation program under grant agreement No 824135 (SOLARNET).
	This work was supported by the Italian MIUR-PRIN grant 2017 ''Circumterrestrial Environment: Impact of Sun--Earth Interaction'' (grant 2017APKP7T), by the German Federal Ministry of Education and Research (Project No. 01LG1909C), and by the BK21 plus program through the National Research Foundation (NRF) funded by the Ministry of Education of Korea.
	This research has made use of NASA's Astrophysics Data System.
\end{acknowledgements}

\bibliographystyle{aa}
\bibliography{_biblio1} 

\appendix
\section{Comparison between reconstructed TSI with RP data and various TSI series}
\label{sec:appendixcomparison}
\begin{figure*}
	\centering
	\begin{minipage}{0.77\linewidth}
		\includegraphics[width=1\linewidth]{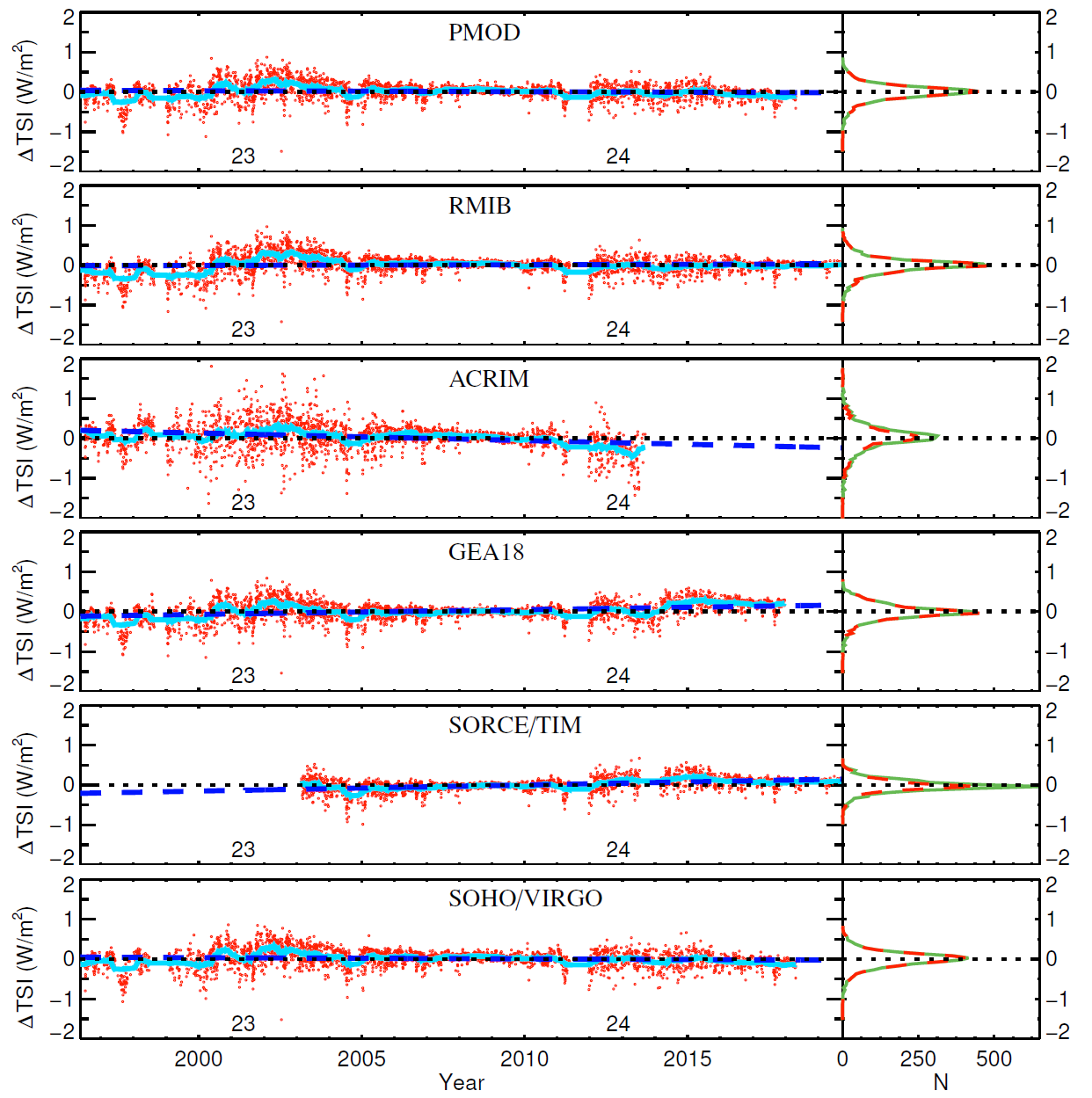}
	\end{minipage}
	\caption{\textit{Left: }
	Difference between the various TSI reference series and the TSI calculated from RP Ca~II~K data (red circles for daily values and light blue curve for 81-day running mean values) using the respective TSI reference series as a function of time. The TSI reference series used are PMOD, RMIB, ACRIM, GEA18, SORCE/TIM, and SOHO/VIRGO. Also shown is a linear fit to the residuals (dark blue dashed line). The horizontal dotted black line denotes the zero level difference. The numbers in the lower part of the panels denote the solar cycle number. \textit{Right: }Distributions of the residuals in bins of 0.05 Wm$^{-2}$ for the series shown in the left panels (red dashed). Also shown (in green) are the distributions normalised to the same number of days, namely that in the PMOD record. }
	\label{fig:psptresiduals2}
\end{figure*}

\begin{figure*}
	\centering
	\begin{minipage}{0.77\linewidth}
		\includegraphics[width=1\linewidth]{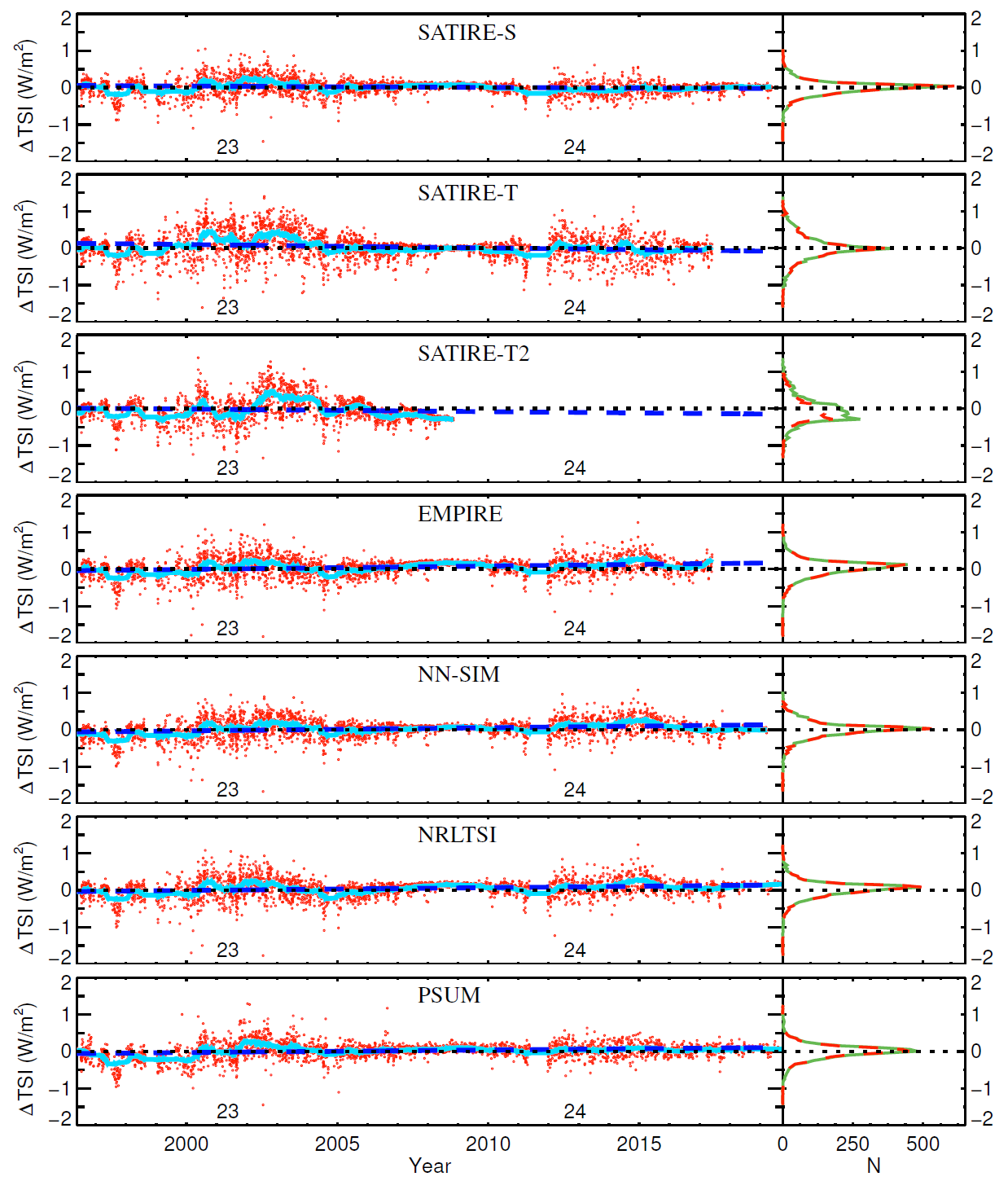}
	\end{minipage}
	\caption{Same as Fig. \ref{fig:psptresiduals2} but showing the differences to the SATIRE-S, SATIRE-T, SATIRE-T2, EMPIRE, NN-SIM, NRLTSI, and PSUM reconstructed TSI series.}
	\label{fig:psptresiduals3}
\end{figure*}

Figure~\ref{fig:psptresiduals2} and \ref{fig:psptresiduals3} show the difference between our reconstructed TSI with RP data and the various respective reference TSI series.
The reconstructed TSI agrees quite well with the various TSI reference series.
The trend of the residuals is generally flat or marginally decreasing,
although for SORCE/TIM and the four empirical models (EMPIRE, NN-SIM,  NRLTSI and PSUM), it is slightly increasing.
We note that PSUM used exactly the same Ca~II~K data for their (empirical) reconstruction. 
The trend is somewhat stronger when we compare to SORCE/TIM, but this could also be, at least partly, due to the shorter period covered by this record.
The  marginally decreasing (increasing) trend means that our reconstructed TSI shows a marginally stronger (weaker) decline with time than the TSI series employed for comparison.
Given that the differences are very small, all results appear consistent with each other. 
Only the ACRIM case is clearly different, with a strong decreasing trend in the residuals.

\section{Effect of bandwidth on the contribution of sunspots to faculae filling factors}
\label{sec:bandwidth_sunspot}
\begin{figure*}
\centering
	\includegraphics[width=1\linewidth]{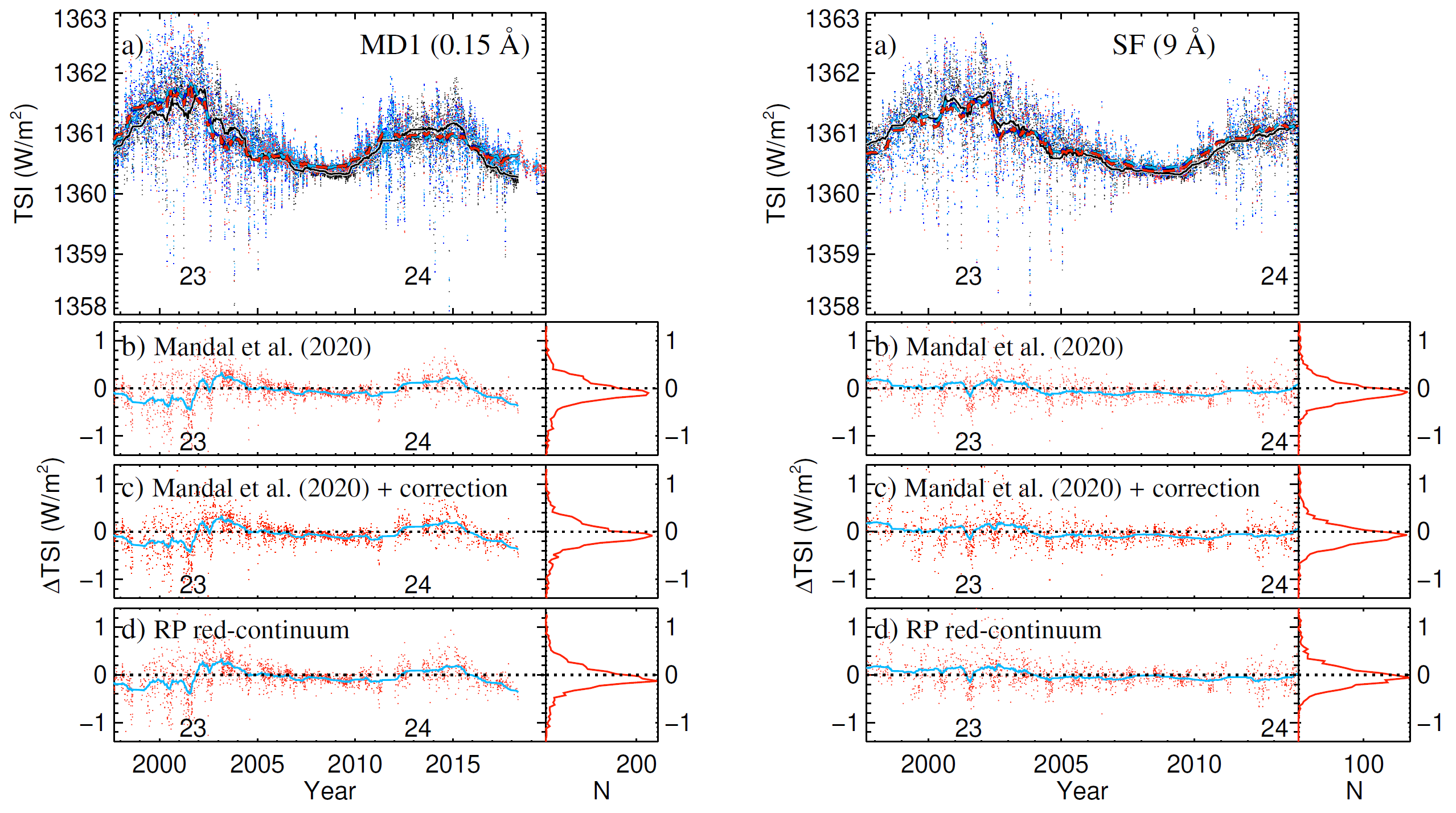}
	\caption{TSI reconstructed with MD1 (left panels) and SF (right panels) Ca~II~K data as a function of time (a) and difference of TSI reconstructions to the PMOD TSI composite (b--d) when using different approaches to account for the sunspots: using the \cite{mandal_sunspot_2020} series to get the sunspot filling factors (light blue in panel a and red in panel b); using the \cite{mandal_sunspot_2020} series to get the sunspot filling factors which were also subtracted from those of faculae (blue in panel a and red in panel c); using full-disc Rome/PSPT red continuum data (red in panel a and d) to get the sunspot filling factors (see Sect. \ref{sec:sunspotcontribution} for more information). The differences are shown only for the common days in all series. Also shown in panel a is the PMOD TSI composite (black). Thick lines show 81-day running mean values. The right part of the lower panels show the distributions of the differences in bins of 0.05 Wm$^{-2}$.}
	\label{fig:difsunspotdif_difarchives}
\end{figure*}

\begin{figure*}
	\centering
	\includegraphics[width=1\linewidth]{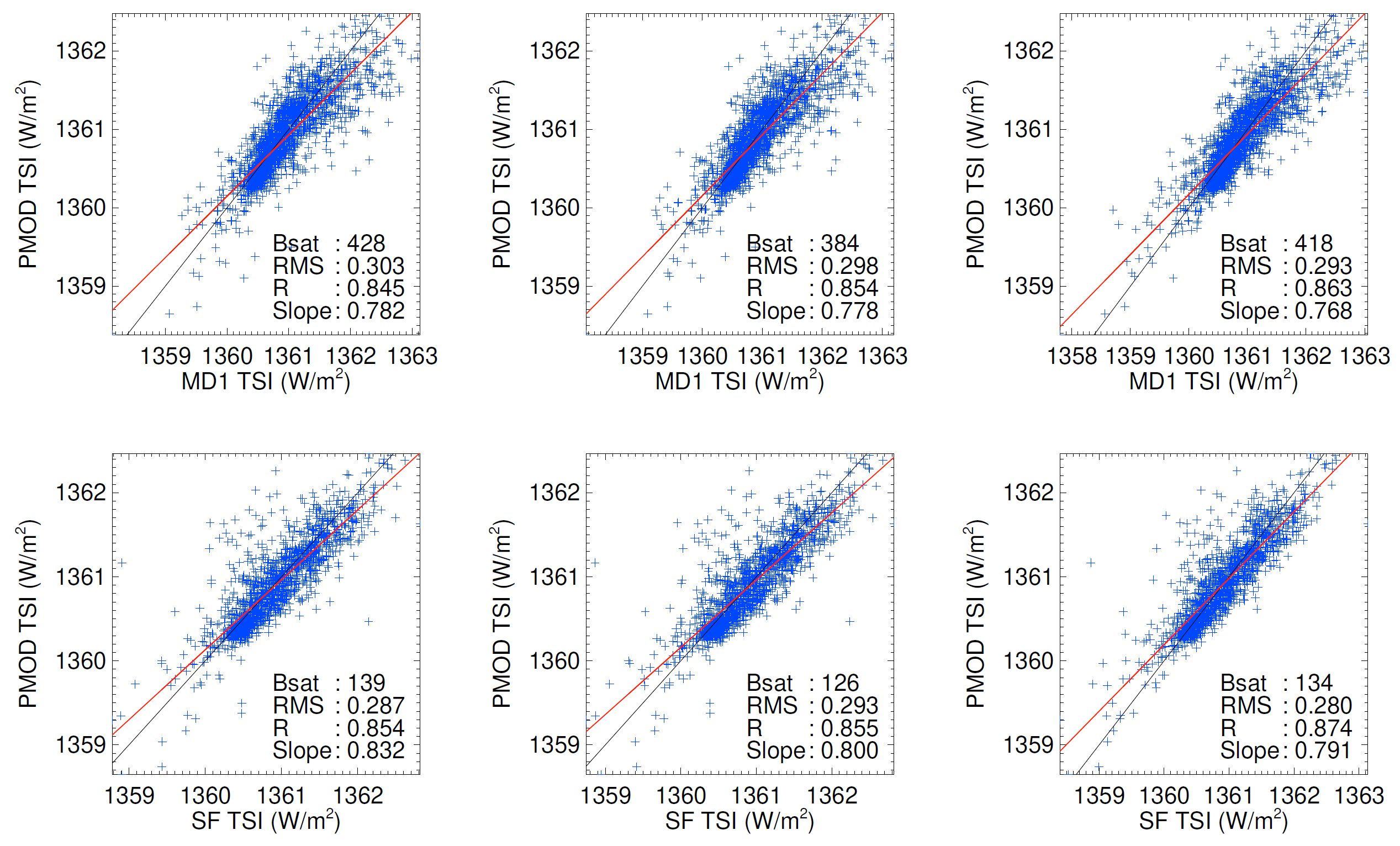}
	\caption{TSI reconstructed from MD1 (top row) and SF (bottom row) Ca~II~K images versus the PMOD TSI composite. \textit{Left:} sunspot information taken from \cite{mandal_sunspot_2020} without any correction to the faculae filling factors due to sunspots; \textit{Middle:} sunspot information taken from \cite{mandal_sunspot_2020} and the sunspot filling factors are subtracted from the facular ones; \textit{Right:} the sunspot information is derived from the full-disc RP red-continuum images. The red lines show linear fits to the data, while the black lines have a slope of unity. Also listed in each panel are the B$_{\mathrm{sat}}$ (in G), RMS difference (in W/m$^{2}$), the linear correlation coefficient, and the slope of the linear fit.}
	\label{fig:scatterdifsunspots_difarchives}
\end{figure*}

To understand whether the contribution of sunspots to the derived facular filling factors is affected by the bandwidth of the Ca~II~K observations, we repeated the analysis presented in Sect. \ref{sec:sunspotcontribution} for the datasets taken with different filters.
Figures~\ref{fig:difsunspotdif_difarchives} and \ref{fig:scatterdifsunspots_difarchives} show the results for the MD1 and SF datasets, which are the two extreme cases of bandwidth analysed here (narrowest and broadest, respectively).
We note that $B_{\mathrm{sat}}$ for the three TSI reconstructions from MD1 and SF Ca~II~K data used for this test was determined by considering the same period, which is limited to the period after September 1997 as dictated by the availability of red-continuum Rome/PSPT observations. This differs to the results presented in the main text with MD1 and SF where the entire period since 1978 and 1992, respectively, was used to set $B_{\mathrm{sat}}$.
For all archives, we report a marginal improvement in the linear correlation and the RMS differences between the reconstructed TSI and the PMOD TSI series when the actual locations of sunspots are used to correct the facular filling factors compared to the case that they are not corrected. 
However, the improvement is rather minor, as already reported for RP data in Sect. \ref{sec:sunspotcontribution}.
This reinforces our conclusion that the information on the sunspot filling factors can be determined from an independent series, such as that of \cite{mandal_sunspot_2020}, without applying any correction to the facular filling factors.
As mentioned before, this is an important result since most historical datasets
do not have co-temporal white light or continuum observations to provide the information on sunspots.

\end{document}